\documentclass[sigconf]{acmart}

\usepackage{fontawesome5}
\usepackage{xcolor}
\usepackage{soul}
\usepackage{wrapfig}
\usepackage{booktabs}
\usepackage{tabularx}
\usepackage{enumitem}
\usepackage{verbatim}
\usepackage{syntax}
\usepackage{parcolumns}
\usepackage{listings}
\usepackage[frozencache,cachedir=.]{minted}
\usepackage[linesnumbered,ruled,vlined]{algorithm2e}
\usepackage{algorithmicx}
\usepackage{algpseudocode}
\usepackage{amsmath}
\usepackage{caption}

\newcommand{\pluto}{\textsc{Pluto}}

\newcommand{\new}[1]{\textcolor{black}{#1}}

\DeclareCaptionType{InfoBox}

\definecolor{explicitBlue}{HTML}{095A94}
\definecolor{customCodeOrange}{HTML}{AB4C0C}
\definecolor{customGreen}{HTML}{54AA54}
\definecolor{customGold}{HTML}{FFD601}

\newcommand{\schemaPrimary}[1]{\texttt{\color{explicitBlue}{#1}}}
\newcommand{\schemaSecondary}[1]{\texttt{\color{customCodeOrange}{#1}}}

\newenvironment{tightItemize}{\begin{itemize}[leftmargin=.15in]\itemsep
-2.1pt}{\end{itemize}}

\newcommand{\key}[1]{\fcolorbox{gray!50}{gray!10}{{\small{#1}}}}

\newcommand{\annotation}[1]{\fcolorbox{customGold}{gray!05}{{\small{#1}}}}

\AtBeginDocument{%
  \providecommand\BibTeX{{%
    \normalfont B\kern-0.5em{\scshape i\kern-0.25em b}\kern-0.8em\TeX}}}

\setcopyright{acmlicensed}
\copyrightyear{2025}
\acmYear{2025}
\setcopyright{cc}
\setcctype{by}
\acmConference[IUI '25]{30th International Conference on Intelligent User Interfaces}{March 24--27, 2025}{Cagliari, Italy}
\acmBooktitle{30th International Conference on Intelligent User Interfaces (IUI '25), March 24--27, 2025, Cagliari, Italy}\acmDOI{10.1145/3708359.3712122}
\acmISBN{979-8-4007-1306-4/25/03}

\begin{document}


\title{\pluto: Authoring Semantically Aligned Text and Charts\\ for Data-Driven Communication}

\author{Arjun Srinivasan}
\affiliation{%
  \institution{Tableau Research}
  \city{Seattle}
  \state{Washington}
  \country{USA}
}
\email{arjunsrinivasan@tableau.com}

\author{Vidya Setlur}
\affiliation{%
  \institution{Tableau Research}
  \city{Palo Alto}
  \state{California}
  \country{USA}
}
\email{vsetlur@tableau.com}

\author{Arvind Satyanarayan}
\affiliation{%
  \institution{Massachusetts Institute of Technology}
  \city{Cambridge}
  \state{Massachusetts}
  \country{USA}
}
\email{arvindsatya@mit.edu}

\renewcommand{\shortauthors}{Srinivasan, et al.}


\begin{abstract}

Textual content (including titles, annotations, and captions) plays a central role in helping readers understand a visualization by emphasizing, contextualizing, or summarizing the depicted data. Yet, existing visualization tools provide limited support for jointly authoring the two modalities of text and visuals such that both convey semantically-rich information and are cohesively integrated. In response, we introduce \pluto, a mixed-initiative authoring system that uses features of a chart's construction (e.g., visual encodings) as well as any textual descriptions a user may have drafted to make suggestions about the content and presentation of the two modalities. For instance, a user can begin to type out a description and interactively brush a region of interest in the chart, and \pluto~will generate a relevant auto-completion of the sentence.
Similarly, based on a written description, \pluto~may suggest lifting a sentence out as an annotation or the visualization's title, or may suggest applying a data transformation (e.g., sort) to better align the two modalities. A preliminary user study revealed that \pluto's recommendations were particularly useful for bootstrapping the authoring process and helped identify different strategies participants adopt when jointly authoring text and charts. Based on study feedback, we discuss design implications for integrating interactive verification features between charts and text, offering control over text verbosity and tone, and enhancing the bidirectional flow in unified text and chart authoring tools.

\end{abstract}



\begin{teaserfigure}
  \includegraphics[width=\textwidth]{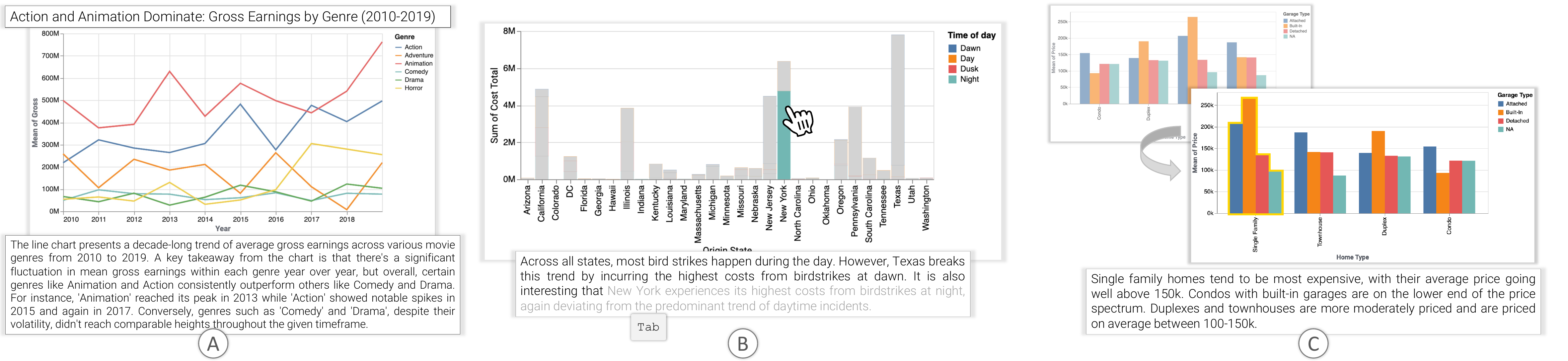}
  \caption{Examples of text and chart suggestions in \pluto. (A) Coherent title and description that are auto-generated based on the chart. (B) Sentence completion is suggested based on multimodal input from the preceding text and a selection on the chart. (C) The chart is sorted and annotated to enhance coherence with the description.}
  \Description[Three examples of text and chart suggestions in Pluto.]{The first example shows a case where the system generates a title and description for a multi-series line chart. The second example shows how the system can complete a sentence in the description when users click on a mark in the chart. The third example illustrates the system's ability to suggest chart design changes, in this case, to short and annotate a chart based on text in the description.}
  \label{fig:teaser}
  \vspace{1em}
\end{teaserfigure}


\maketitle

\section{Introduction}

Research has shown that, perhaps paradoxically, text plays an important role in how readers interpret the data depicted by a visualization. These descriptions can explain how a chart is constructed, summarize statistical features (e.g., the minimum and maximum values), describe cognitive and perceptual phenomena (e.g., complex trends and patterns), or offer broader contextual and domain-specific explanations~\cite{lundgard2021accessible}.
When well-authored, textual descriptions can helpfully reinforce visual features of the chart (e.g., high prominence characteristics) to reduce a reader's cognitive load while interpreting the chart~\cite{kim2021towards}\,---\, indeed, in such cases, readers favor heavily annotated charts over simply charts or textual descriptions alone~\cite{stokes2022striking}. However, poorly written text can also slant the takeaway message~\cite{kong2018frames} in ways that impact readers' trust and recall~\cite{kong2019trust}.

Existing tools, however, provide almost no support for authoring text alongside the visualization. At best, a nascent body of work has begun to explore automated methods for generating titles~\cite{liu2023autotitle} and descriptions~\cite{choi2022intentable,obeid2020chart,hsu2024scicapenter} or, for pre-authored visualizations and textual descriptions, aligning emphasis~\cite{kim2023emphasischecker}, or constructing links and cross-references~\cite{latif2021kori}. While valuable, these approaches leave unexplored a rich design space for \emph{concurrently} authoring the two modalities together\,---\, that is, how might interactive mechanisms help users write a textual description while designing a chart; conversely, how might users update a chart's design to better reflect a textual description; and, how might a user iterate between the two modalities when authoring artifacts for data-driven communication?

To address this gap, we introduce \pluto\footnote{Inspired by the celestial narrative where Pluto represents an entity once marginalized in astronomical classification, this work seeks to underscore the often-overlooked importance of text in data visualization. The name also aligns with the tradition of naming advances in visualization after celestial bodies, such as Polaris~\cite{polaris} and Vega~\cite{vega2014visualization}.}, a novel mixed-initiative tool designed to enable authors to craft semantically rich textual content for a variety of popular charts, including bar charts, histograms, line charts, and scatterplots.
\pluto~facilitates an interactive composition process, allowing authors to leverage GPT for automatically generating chart descriptions (Figure~\ref{fig:teaser}A), use direct manipulation-based interactions with the chart to scope and guide text generation (Figure~\ref{fig:teaser}B), and even update the chart's design based on the composed text (Figure~\ref{fig:teaser}C). To operationalize this interactive authoring experience, we model a conceptual schema that captures key components of the chart (i.e., data, specification, selections, visual embellishments) and text (i.e., title, description, textual annotations along with the semantic information they convey~\cite{lundgard2021accessible}).
We leverage this schema as part of a pipeline that blends heuristics and large language models (LLMs) to generate authoring recommendations in \pluto.

Through a preliminary user evaluation, we explore the utility of \pluto{} as a tool for helping authors create semantically relevant content when integrating text with the corresponding chart that it describes. We observe that the recommendations collectively afford a breadth of authoring workflows along a spectrum of using automatically generated text to leveraging recommendations only to augment manually authored content. Participant feedback also suggests that certain forms of assistance---such as generating text based on chart selections or suggesting chart design changes based on entered text---are deemed more useful than automated text generation and text editing suggestions. Our findings contribute to a growing body of literature on the integration of textual and visual elements for data-driven communication, offering insights into how future visualization authoring tools can better leverage the semantics and functional role of text with visualizations.
\section{Related Work}
Our work builds on several lines of research: exploring the role of text with visualizations, visualization and text systems, and image and text authoring interfaces.

\subsection{The role of text with visualizations}
The interplay between text and visual elements in data visualization has been a significant area of interest with increased advocacy for treating text as co-equal to visualization~\cite{stokesgive, lundgard2021accessible}. Kim et al.~\cite{kim2021towards} conducted a study to understand how readers integrate charts and captions in line charts. The study findings indicated that when both the chart and text emphasize the same prominent features, readers take away insights from both modalities. Their research underscores the importance of coherence between visual and textual elements and how external context provided by captions can enhance the reader's comprehension of the chart's message. Building on these insights, Lundgard and Satyanarayan~\cite{lundgard2021accessible} proposed a four-level model for content conveyed by natural language descriptions of visualizations. Their model delineates semantic content into four distinct levels: elemental and encoded properties (Level 1), statistical concepts (Level 2), perceptual and cognitive phenomena (Level 3), and contextual insights (Level 4).

Focusing on the role of textual annotations in visualization, Stokes et al.~\cite{stokes2022striking} observed that readers favored heavily annotated charts over less annotated charts or text alone. This preference highlights the added value of textual annotations in aiding data interpretation, with specific emphasis on how different types of semantic content impact the takeaways drawn by readers. Further contributions by Quadri et al.~\cite{quadri2024you} and Fan et al.~\cite{fan2024understanding} explored high-level visualization comprehension and the impact of text details and spatial autocorrelation on reader takeaways in thematic maps. These studies collectively underline the critical role of textual elements in shaping viewer perceptions, understanding, and accessibility of visual data. Ottley et al.~\cite{ottley2019curious} and Stokes et al.~\cite{stokes2023role} have also contributed to this body of research, focusing on how annotations influence perceptions of bias and predictions, reinforcing the multifaceted impact of text on visual data interpretation.

Our work further explores how text and charts can be better aligned with one another by offering a mixed-initiative authoring interface. Specifically, \pluto~allows leveraging both direct manipulation interactions and user-drafted text to generate recommendations for communicative text and chart design. Furthermore, \pluto's text recommendations explicitly incorporate Lundgard and Satyanarayan's model~\cite{lundgard2021accessible} for semantic information conveyed by visualization descriptions.
In doing so, the system ensures that the generated text has good semantic coverage and structure (e.g., generated descriptions start by conveying the chart's encodings and then list high-level trends) and is appropriate for the intended communicative use (e.g., the semantic information conveyed by titles is different from descriptions accompanying a chart or annotations on the chart).

\subsection{Visualization and text systems}

The integration of visualization and text has led to the development of various systems designed to facilitate the creation, interpretation, and enhancement of data visualizations with textual elements. He et al.~\cite{he2024leveraging} surveyed the leveraging of large models for crafting narrative visualizations, highlighting the potential of AI in supporting the narrative aspect of data visualization. This is complemented by AutoTitle, an interactive title generator for visualizations~\cite{liu2023autotitle}, and Vistext, a benchmark for semantically rich chart captioning~\cite{tang2023vistext}. VizFlow demonstrates the effectiveness of facilitating author-reader interaction by dynamically connecting text segments to corresponding chart elements to help enrich the storytelling experience~\cite{sultanum2021}. This body of research highlights the need for tools to support more nuanced integration of text and visualization.

Supporting the co-authoring of text and charts, Latif et al. introduced Kori~\cite{latif2021kori}, an interactive system for synthesizing text and charts in data documents, emphasizing the seamless integration of visual and textual data for enhanced communication.
\new{CrossData~\cite{chen2022crossdata} presents an interactive coupling between text and data in documents, enabling actions based on the document text and adjusting data values in the text through direct manipulation on the chart.
Such systems illustrate the potential for the bidirectional linking between text and charts to assist rich authoring of data-driven narratives.
}
Furthermore, systems like EmphasisChecker~\cite{kim2023emphasischecker}, Intentable~\cite{choi2022intentable}, Chart-to-text~\cite{obeid2020chart}, DataDive~\cite{kim2024datadive},
\new{InkSight~\cite{lin2023inksight}},
and FigurA11y~\cite{singh2024figura11y} focus on guiding chart and caption creation, supporting readers' contextualization of statistical statements, and assisting in writing scientific alt text. Recent work like SciCapenter supports the composition of scientific figure captions using AI-generated content and quality ratings \cite{hsu2024scicapenter}.
DataTales~\cite{sultanum2023datatales} is another example of a recent system using a large language model for authoring data-driven articles, indicating the growing interest in AI-assisted data storytelling.
These systems collectively illustrate the expanding scope of text integration into visualization, from enhancing data document creation to improving accessibility and data-driven communication. Reviewing the aforementioned tools and the use of generative AI for visualization more broadly, Basole and Major~\cite{basole2024generative} discuss how generative AI methods and tools offer creativity assistance and automation within the visualization workflow, specifically highlighting a shift towards ``human-led AI-assisted'' paradigms, where generative AI not only augments the creative process but also becomes a co-creator.

Aligned with this paradigm shift, \pluto~adopts a mixed-initiative approach that leverages the capabilities of generative AI to help create semantic alignment between the chart and its corresponding text for effective data-driven communication.
However, \pluto~differs from existing chart-and-text authoring tools in three significant ways.
First, going beyond existing systems that primarily leverage unimodal information from the chart to generate text, \pluto~supports multimodal authoring combining information from both the chart (including any direct interactions with marks) and user-drafted text.
Furthermore, unlike prior tools that focus exclusively on generating complete descriptions/captions or titles, \pluto's recommendations can be leveraged in flexible ways to author not only titles and descriptions but also more fine-grained annotations and sentence completions. Second, while existing tools primarily recommend text for a given chart, \pluto's recommendations are bidirectional.
Specifically, the system suggests chart design changes like sorting or adding embellishments based on the authored text, resulting in artifacts that more clearly communicate takeaways via a combination of text and charts. Lastly, unlike existing tools that primarily rely on pre-trained knowledge in generative AI models, \pluto's recommendations are grounded in a theoretical research-based model of semantic information conveyed in visualization text~\cite{lundgard2021accessible}, ensuring the generated text covers the appropriate level of detail and is effective for communication \emph{alongside} the chart.
\begin{figure*}[t!]
    \centering
    \includegraphics[width=\linewidth]{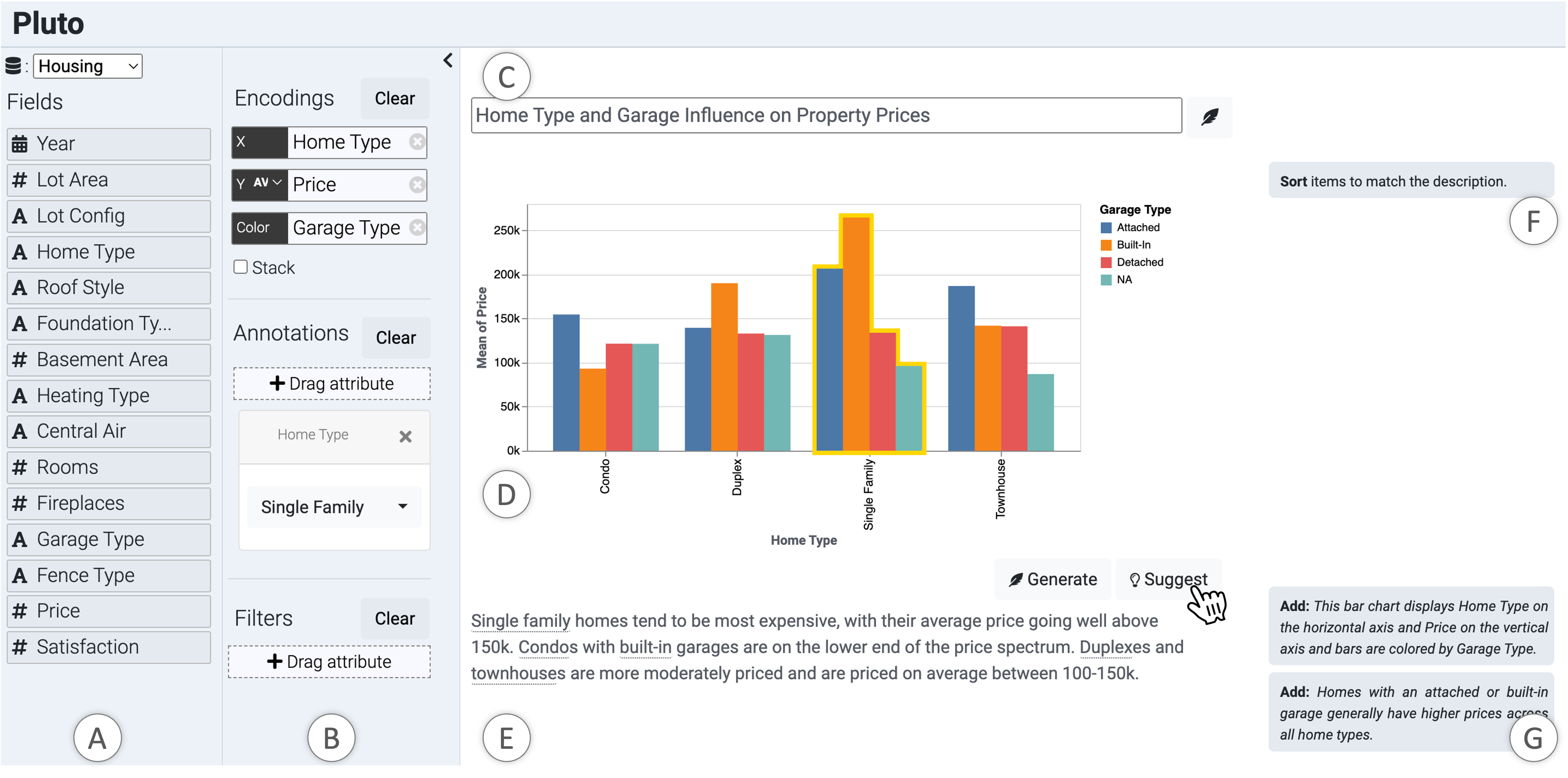}
    \caption{\pluto's user interface. The key components include a data panel (A), chart editor (B), chart title (C), main chart canvas (D), and a chart description (E).
    Here, the user has manually entered a description and clicked the {\small{\faIcon[regular]{lightbulb}}} \textbf{Suggest} button to get ideas on improving the chart and text for communication purposes.
    This results in the system suggesting a title and adding a highlight annotation for \annotation{\textit{Single Family}} homes, while also generating a chart design recommendation (F) and a set of description editing recommendations (G).}
    \Description[Pluto's user interface.]{From left to right, the system contains of: 1) a data pane showing the available attributes, 2) a pane to specify encodings, annotations, and filters to create a chart, 3) the chart along with text boxes for the title and description above and below it, respectively, and 4) a right panel where the system presents recommendations to edit the chart and description.}
    \label{fig:interface}
\end{figure*}

\section{Design}

The central idea of our work is exploring a unified visualization system for authoring well-integrated charts and text.
Designing such a system, however, requires considering several open questions about the type of assistance the system should provide, and when and how system suggestions should be surfaced.

In exploring the chart-and-text authoring experience, several questions arise that warrant exploration. We must first discern which chart elements, ranging from axis labels and ticks to titles, descriptions, and annotations, necessitate the most authoring support. Additionally, we would need to determine the appropriate level of system assistance—whether to generate entire descriptions, fill in partially written text, or refine user-authored drafts—and how this assistance might vary with different types of text. Given the non-mutual exclusivity of text types, such as descriptions influencing titles, we must also consider if and how the system should sequence its suggestions. The timing of these suggestions is another critical factor: should they be offered immediately following the creation of a chart or once the user has initiated the authoring process? Deciding whether these suggestions should be proactive or solicited on-demand, along with the specific user actions that should trigger them, is also a direction worth considering. Lastly, we must explore the potential for a synergistic relationship between the chart and text, i.e., how interactions with each can be leveraged to enhance the other and what mechanisms would facilitate this interplay.

\subsection{Design Goals}
\label{sec:design-goals}

With the aforementioned considerations in mind, we iteratively compiled a list of design goals to guide our system's development.
These goals were informed by prior research and systems focusing on authoring text for visualizations (e.g.,~\cite{kim2023emphasischecker,latif2021kori,liu2023autotitle,he2024leveraging,tang2023vistext,singh2024figura11y}), general principles of mixed-initiative user interfaces~\cite{horvitz1999}, as well as formative interviews with two experts on authoring text and charts for data-driven communication.

The two experts were a practitioner and a researcher who both author and critique text for data visualizations. 
Furthermore, they also regularly interact with end-users in the creation of text and charts.
\new{Both experts voluntarily participated in the interviews and were not financially compensated.
We interviewed each expert twice over a span of three weeks.
Each session lasted for 30-60 minutes.
}
\new{During the first set of interviews, we asked the experts about the key challenges users generally encounter during the authoring of text with charts. Additionally, to guide our design and identify critical features, we also presented an early version of our prototype with a basic set of functionality including generating titles and descriptions for a chart and supporting interactive highlighting of chart elements based on the text.
Based on the initial feedback, we incorporated additional types of recommendations and refined the system design before the second meeting where the experts provided feedback on the overall utility and perceived usability of the different features.
We subsequently developed the final version of \pluto~by iterating on this feedback and leveraging findings from related work on text+chart authoring systems~(e.g.,~\cite{latif2021kori,kim2023emphasischecker,sultanum2023datatales,lin2023inksight,choi2022intentable}).
}

\vspace{.5em}
\noindent\textbf{DG1. Leverage the textual narrative to guide chart design.}
In line with prior work~\cite{stokes2022striking,ottley2019curious,kim2021towards}, the experts also stressed that the text should not only convey the right levels of information but also be well-aligned with the chart for a smooth reading experience.
As text has an inherent narrative flow, we noted that the system should \new{incorporate techniques from prior work on updating chart specification based on narrative text~\cite{wang2022towards,chen2022crossdata,shen2024data} to} inspect the flow of information in the text and leverage it to augment the chart.
This augmentation could involve making data transformation changes (e.g., sorting) or adding annotations to highlight portions of the chart that are emphasized in the text.

\vspace{.5em}
\noindent\textbf{DG2. Support direct manipulation interactions with the chart for text generation.}
Visualizations make it easy to perceive trends in the data and identify points of interest.
Phrasing something visually interesting as text can be challenging, however. For instance, one of the experts noted, ``\textit{sometimes I notice something potentially interesting on the chart and want some quick text to verify what I'm seeing and get ideas for how to talk about it.}'' Given this multimodal nature of charts and text, \new{in line with prior chart-and-text authoring systems (e.g.,~\cite{chen2022crossdata,lin2023inksight})}, we noted that the system should allow leveraging direct interactions with the chart (e.g., brushing a region or mark selection) to generate corresponding text.

\vspace{.5em}
\noindent\textbf{DG3. Provide varying levels of assistance for text authoring.}
Both experts noted that users need different levels of assistance when writing text depending on their goals and experience level.
For instance, novice and intermediate users may need auto-generated text to jump-start their authoring process, whereas domain experts may benefit from fine-tuning suggestions to improve manually written text.
Combining this comment with prior work on text-chart authoring~\cite{stokes2022striking,kim2023emphasischecker}, we noted that the system should not only recommend text for chart authors to add but also recommend editing actions (e.g., reordering sentence) or flag potential factual errors in the text (e.g., incorrectly stated trends).

\vspace{.5em}
\noindent\textbf{DG4. Incorporate context-sensitive recommendations near their relevant targets to facilitate easier interpretation.}
System recommendations in the context of a unified text and chart authoring process could apply to different targets (e.g., chart, title, or description) and focus on either adding new content or editing existing content.
Interpreting this broad set of recommendations can be challenging, however.
For instance, in our early prototypes, we explored listing all recommendations in a side panel, but both experts noted that this was overwhelming and distracted them from the main content.
Iterating on the designs, we noted that for improved usability, the system recommendations should be placed close to the targets they apply to and should also be presented differently (e.g., in-place overlays vs. suggested actions) based on the type of recommendation.

\vspace{.5em}
\noindent\textbf{DG5. Recommendations should be unobtrusive during targeted authoring.}
While the recommendations are designed to help craft cohesive text and charts, there may be instances where the chart authors have clear authoring goals in mind.
In such targeted authoring scenarios, the recommendations should not interfere with the users' flow but still be available on demand if users want ideas for text content or chart design.
Authors should have full control over the final content, however, and should be able to edit/update any suggestions made by the system.
\newline

\noindent{}Note that these goals are not exhaustive or mutually exclusive, nor are they meant to be prescriptive.
For instance, we primarily focus on content suggestions and do not deeply consider operations like formatting as part of the recommendation space. Rather, \textbf{DG1}-\textbf{DG5} are only meant to be an initial set of goals to help ground our design and enable us to develop and test a viable prototype.
\section{Pluto}

Incorporating these design goals, we implemented \pluto~as a prototype system for authoring semantically-aligned text and charts.

\subsection{Example Usage Scenarios}
\label{sec:scenarios}

\begin{figure}[t!]
    \centering
    \includegraphics[width=.5\textwidth]{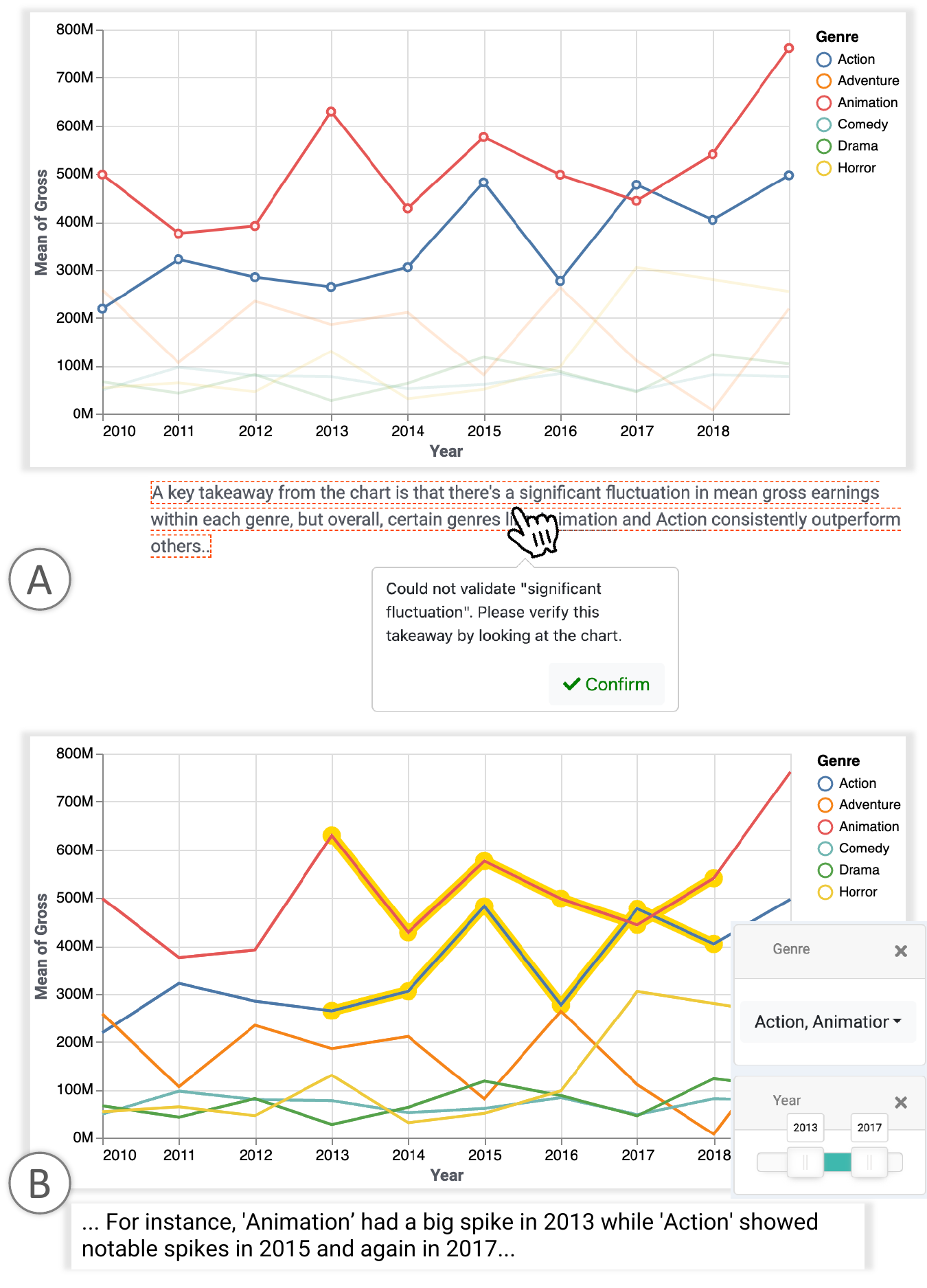}
    \caption{Upon processing a description, \pluto~flags statements that require manual verification (A) and automatically \annotation{annotates} the chart based on data references in the description (B).}
    \Description[Two examples shows Pluto's suggestions for statement verification and chart annotation.]{In the first case, the system highlights a potentially incorrect statement in the user's description of the chart, allowing the user to inspect and verify whether the statement is correct. The second example illustrates how the system annotates (here, by adding a gold stroke) a portion of the chart by inspecting the data value references in the description.}
    \label{fig:scenario-1}
\end{figure}

Figure~\ref{fig:interface} shows \pluto's interface.
Users can drag and drop data fields onto visual encoding channels to create charts.
To underscore a unified experience for authoring charts and text, the system also presents an explicit title and description region just above and below the chart.
Users can also annotate the chart by creating text callouts via a context menu invoked on the chart, or by adding visual embellishments using the chart editor (e.g., borders to highlight marks).
System recommendations are either directly applied to the title, chart, or description or displayed to the right of the chart and description (Figure~\ref{fig:interface}F, G) for authors to review (\textbf{DG4}, \textbf{DG5}).

To illustrate how \pluto's interface and features collectively enable unified authoring of text and charts for data-driven communication, we now describe three vignettes\footnote{These usage scenarios are modeled on examples of how participants used \pluto~during the study described in Section~\ref{sec:study}.}.
These examples are also illustrated in the supplementary video.

\vspace{.5em}\noindent\textbf{Augmenting generated text with built-in safeguards and chart annotations.}
Consider Dinah, an analyst at a movie production company.
Dinah is tasked with summarizing a chart showing movie earnings across genres (Figure~\ref{fig:teaser}A) to share as part of a report her company plans to publish.

Dinah is unsure about how to start her description, so she uses the {\small\faIcon{feather-alt}} \textbf{Generate} feature to bootstrap her authoring process (\textbf{DG3}).
In response, \pluto~inspects the chart and returns a description for Dinah to review (Figure~\ref{fig:teaser}A-bottom).
Dinah peruses the generated text and manually edits it for conciseness.
She then clicks {\small{\faIcon[regular]{lightbulb}}} \textbf{Suggest} to get ideas for using a combination of the description and the chart for better communication.

Analyzing the description, \pluto~makes three changes.
The system flags
description statements with ambiguous takeaways
for review using a dashed border, suggesting that Dinah manually verifies the text with the chart before sharing it with others (Figure~\ref{fig:scenario-1}A) (\textbf{DG3}).
Using \pluto's interactive highlighting feature, Dinah hovers over the statement in the description to see portions of the chart it refers to.
Reflecting on the flagged text, Dinah updates it to remove the modifier ``\textit{significant}'' and make her description more objective for the readers' interpretation.

\pluto~also suggests a title, ``\textit{Action and Animation Dominate: Gross Earnings by Genre (2010-2019)}'' based on both the narrative in the description and the underlying trends in the chart (Figure~\ref{fig:teaser}A-top).

Finally, besides suggestions for the text, \pluto~also adds an annotation to the chart highlighting the key regions the text describes (\textbf{DG1}).
In this case, detecting the emphasis on the \textit{Action} and \textit{Animation} genres and their trends between 2013 and 2017, the system adds a \annotation{gold stroke} around the corresponding lines in the chart (Figure~\ref{fig:scenario-1}B).

Satisfied with her changes based on the system suggestions, Dinah shares the title, chart, and description with her colleagues for review.

\begin{figure}[t!]
    \centering
    \includegraphics[width=\linewidth]{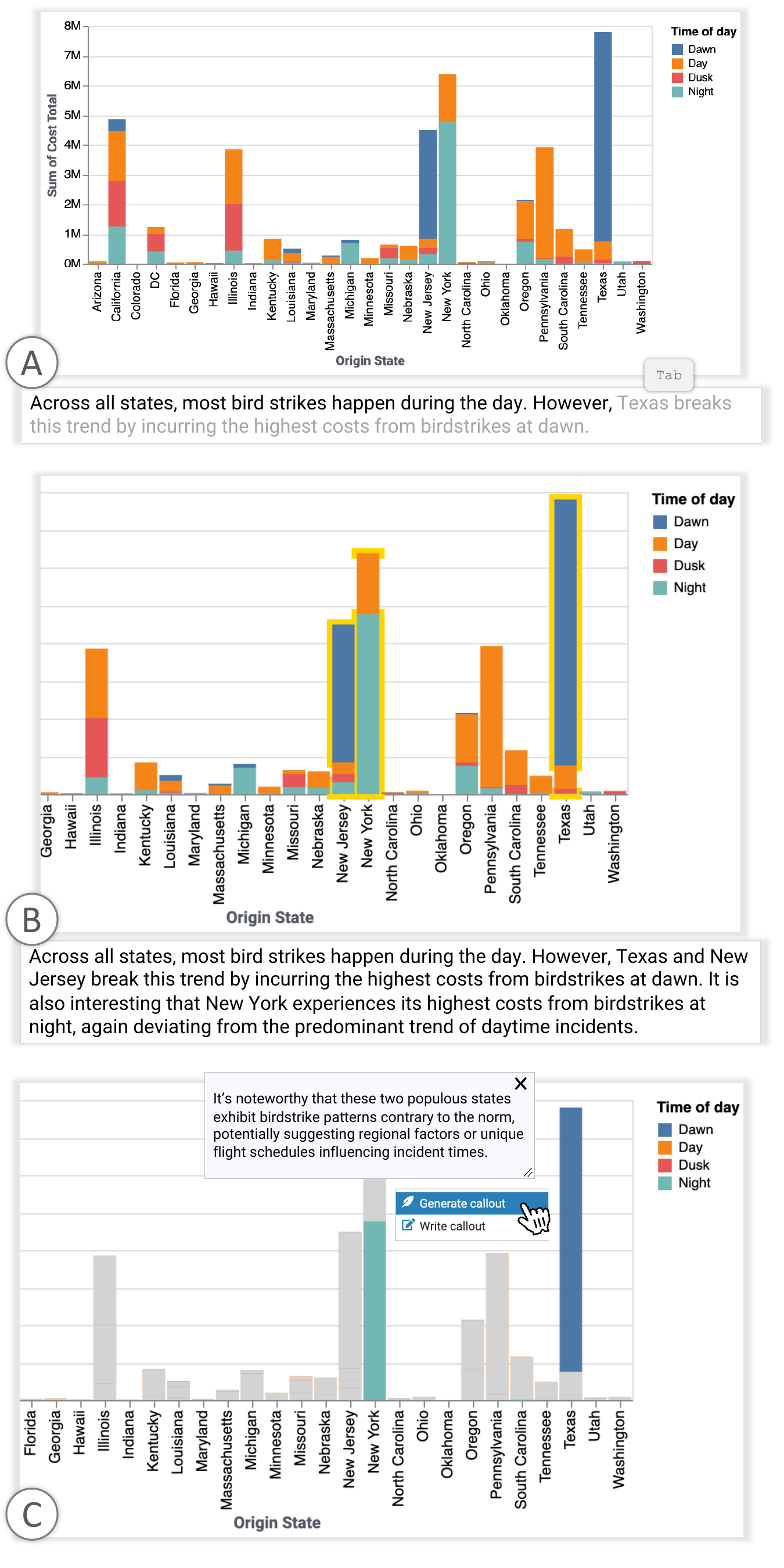}
    \caption{Examples of \pluto's recommendations including an in-place sentence completion (A), \annotation{annotations} based on a chart's description (B), and a text callout generated based on marks selected on a chart (C).}
    \Description[Three examples of Pluto's recommendations.]{In the first case, the system completes the user's sentence when the user presses the tab key on a partially typed statement. In the second example, the system adds a gold stroke to marks on the chart to highlight that they are referenced in the text. In the third example, the user clicks two marks on the chart and asks Pluto to generate a text callout - this results in the system adding a text bubble with content focusing on the selected marks.}
    \label{fig:scenario-2}
\end{figure}

\vspace{.5em}
\noindent\textbf{Steering text competition through data-driven narratives and multimodal input.}
Imagine Ronnie, a financial analyst, writing a report on the monetary impact of bird strikes across the US based on the chart shown in Figure~\ref{fig:scenario-2}A.

Ronnie notices that most states, with the exception of Texas and New Jersey, have tall orange bars corresponding to costs incurred by bird strikes during the day.
Noting this observation, Ronnie types, ``\textit{Across all states, most bird strikes happen during the day.}''
Wanting to emphasize the exception of Texas and New Jersey, Ronnie types ``\textit{However, }'' and presses the \key{Tab} key to ask the system to finish the sentence.
Parsing the preceding sentence and the data trends from the chart, \pluto~generates the completion ``\textit{Texas breaks this trend by incurring the highest costs from birdstrikes at dawn.}'' (Figure~\ref{fig:scenario-2}A)
Ronnie accepts this completion but edits it to include New Jersey.

Next, to emphasize the high cost incurred by incidents in New York at night, Ronnie types ``\textit{It is also interesting that}," clicks on the teal bar showing the total cost for \textit{New York} at \textit{Night}, and again invokes a sentence completion.
Using the multimodal input from the chart selection and the existing description text (\textbf{DG2}), \pluto~suggests the text ``\textit{New York experiences its highest costs from birdstrikes at night, again deviating from the predominant trend of daytime incidents.}'' (Figure~\ref{fig:teaser}B)

Content with his description, Ronnie uses {\small{\faIcon[regular]{lightbulb}}} \textbf{Suggest} to see how he can further improve his text and the chart.
\pluto~processes the description and adds a \annotation{stroke} to visually highlight the states \textit{Texas}, \textit{New York}, and \textit{New Jersey} and the times \textit{Dawn} and \textit{Night} based on their high data values and the emphasis in the description (Figure~\ref{fig:scenario-2}B) (\textbf{DG1}).
Seeing this annotation gives Ronnie an idea to explicitly call out the striking differences in values between Texas and New York.
He selects the two tall bars within Texas and New York and uses {\small{\faIcon[regular]{feather-alt}}} \textbf{Generate Callout} to create a textual annotation directly overlaid onto the chart (Figure~\ref{fig:scenario-2}C) (\textbf{DG2}).
Manually refining the generated callout and visual embellishments (\textbf{DG5}), Ronnie saves the annotated chart and description for his report.

\vspace{.5em}
\noindent\textbf{Guiding manual authoring via system recommendations.}
Imagine Fifi, a realtor who is using the grouped bar chart shown in Figure~\ref{fig:interface}D to author an email blast on house pricing trends for her clients.
Analyzing the chart, Fifi manually writes a description with three sentences, shown in Figure~\ref{fig:interface}E.
With this initial text, she invokes the {\small{\faIcon[regular]{lightbulb}}} \textbf{Suggest} feature to see how she can improve her text and chart for communication.

Parsing the description, \pluto~detects that it lacks a summary of the chart's encodings and also detects that there is no higher-level statement encompassing a trend across multiple home types.
Translating these into recommendations, \pluto~suggests adding a brief statement about the chart's layout and a statement talking about the general impact of garage types across house types, respectively (Figure~\ref{fig:interface}G) (\textbf{DG3}).
Acknowledging these might be useful as overview statements for her readers, Fifi previews what her description would read like with the suggested text by hovering on the recommendations and subsequently accepting them.
As with the other examples, \pluto~also suggests a title (\textit{Home Type and Garage Influence on Property Prices}), but Fifi finds this too formal and manually adjusts it to make it more catchy: ``\textit{Can I afford both a car and a home?: The Influence of Garage Type on Property Prices}.''

\begin{figure}[t!]
    \centering
    \input{listings-and-algos/definitions}
    \caption{Conceptual schema representing the key text and chart elements in \pluto's interface.}
    \Description[Conceptual schema representing the key text and chart elements in Pluto.]{Conceptual schema representing the key text and chart elements in Pluto}
    \label{fig:schema}
\end{figure}

Besides the text suggestions, \pluto~also detects that the description emphasizes the home types with highest and lowest values, whereas the home types in the chart are sorted alphabetically.
To resolve this disparity, the system provides a chart design recommendation to sort the home types by price ranging from the highest to lowest (Figure~\ref{fig:interface}F) (\textbf{DG1}).
Fifi accepts this recommendation as it can give her clients a glanceable summary of some key takeaways in her text (Figure~\ref{fig:teaser}C).

\subsection{Conceptual Model}

To enable the aforementioned workflows and recommendations, we model the various components across the text and the chart in \pluto~as a \emph{conceptual schema} summarized in Figure~\ref{fig:schema}.
We use this schema in the subsequent sections to detail how the system tracks user input and generates recommendations.

Specifically, a \schemaPrimary{Chart} is represented by mapping \schemaPrimary{Data} onto specific visual encodings (in this case, a Vega-Lite \schemaPrimary{Specification}~\cite{satyanarayan2016vega}).
The \schemaPrimary{ActiveSelection} enumerates the data items that have been selected through direct manipulation interaction (e.g., in Figures~\ref{fig:teaser}B,~\ref{fig:scenario-2}C).
Additionally, the chart can also have one or more \schemaPrimary{Annotations}.
These can be textual comments on the chart, visual embellishments applied to individual marks (e.g., Figures~\ref{fig:interface}D,~\ref{fig:scenario-1}B), or overlays like a regression line on a scatterplot or a line marking the average value across all bars in a bar chart.

The chart's \schemaPrimary{Title} is \schemaPrimary{Text} that may include references to \schemaPrimary{DataItems} (e.g., \textit{Genre}: [\textit{Action}, \textit{Animation}] in Figure~\ref{fig:teaser}A).

\begin{figure}[t!]
    \centering    
    \includegraphics[width=\linewidth]{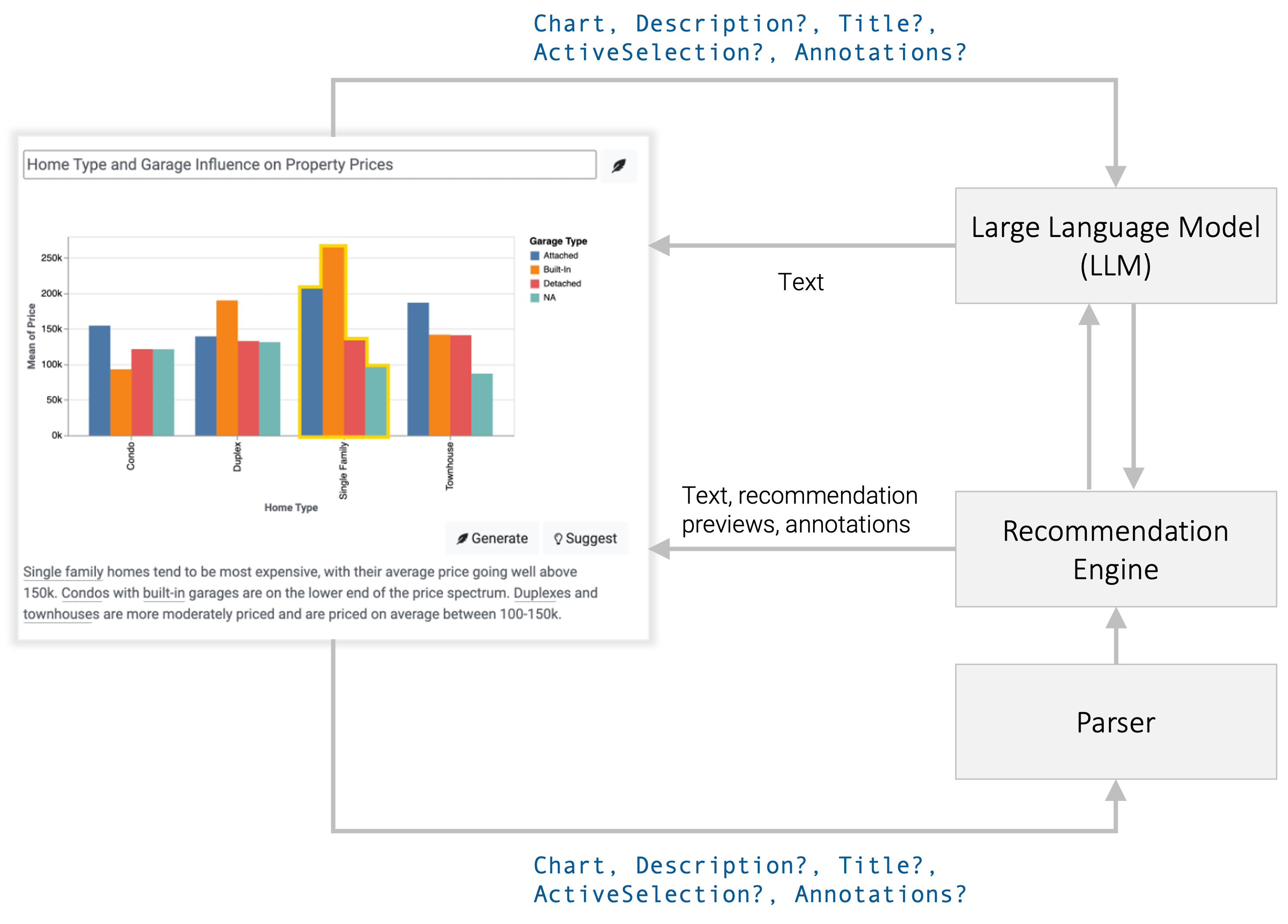}
    \caption{\pluto's system architecture overview}
    \Description[Pluto's system architecture.]{Given the active chart and text from the interface, Pluto uses a heuristic parser process the information and to generate recommendations. When providing recommendations involving text suggestions, the system also uses an LLM in parallel to generate the recommended text.}
    \label{fig:architecture}
\end{figure}

The \schemaPrimary{Description} is represented as a collection of \schemaPrimary{Statements}.
Each statement maps to one of the four semantic statement types proposed by Lundgard and Satyanarayan~\cite{lundgard2021accessible}---namely, \schemaSecondary{encoding}, \schemaSecondary{perceptual-trend}, \schemaSecondary{data-fact}, \schemaSecondary{domain-specific}, or  \schemaSecondary{other} (e.g., a statement about the data source for a chart).
Additionally, similar to the title, statements in the description may also contain references to specific \schemaPrimary{DataItems}.

Note that this schema is not exhaustive (e.g., there may be additional types of annotations, statement types, or chart selections) and was primarily designed to operationalize the recommendations in \pluto.

However, we hope that the idea of formalizing not only the chart but also its associated text can inspire future work on grammars and systems for data-driven communication through a \emph{combination} of text and charts.

\subsection{System Overview}

\pluto~is implemented as a web-based application and is developed using Python, HTML/CSS, and JavaScript.
Visualizations in the tool are created using Vega-Lite~\cite{satyanarayan2016vega}.
The system currently supports three encoding channels (\texttt{x}, \texttt{y}, \texttt{color}) and three mark types (\texttt{bar}, \texttt{line}, \texttt{point}).
Collectively, this combination of encoding and mark types enables specifying several visualizations, including single- and multi-series bar charts, line charts, histograms, and scatterplots, covering a breadth of visualizations explored in prior systems~\cite{kim2023emphasischecker,latif2021kori,sultanum2023datatales,kim2024datadive,choi2022intentable,obeid2020chart,liu2020autocaption,hsu2021scicap,alam2023seechart}.

Figure~\ref{fig:architecture} depicts a high-level overview of the system architecture.
Specifically, \pluto~uses a combination of an LLM (GPT-4~\cite{achiam2023gpt}) and a heuristics-based approach for generating suggestions.
Specifically, requests like generating an entire description or a title from a chart are directly fulfilled using the LLM.
In other cases, a custom parser extracts information from the text and chart and also classifies statements in the description based on their semantic levels~\cite{lundgard2021accessible}.
This extracted information is leveraged by a heuristics-based recommendation engine to generate recommendations, including adding/editing text, adding mark annotations, and suggesting chart design changes such as sorting, among others.
In cases where the recommendations involve generated text suggestions, the recommendation engine either uses its built-in templates or interacts with the LLM to pass it the required context for the text generation.
In the subsequent sections, we detail these components and \pluto's recommendation generation process.

\subsection{Text and Chart Parsing}
\label{sec:parser}

The parser extracts a number of features from the text and the chart that are used to determine system recommendations.

\textbf{Text.}
The parser analyzes text in the description, title, and annotations to identify \schemaPrimary{DataItems}.
The system uses a combination of a lexicon- and grammar-based approach adapted from prior natural language interfaces for visualization (e.g.,~\cite{gao2015datatone,setlur2016eviza,narechania2020nl4dv}) to detect data item references.
Specifically, given an input text, the parser extracts a list of N-grams and compares the N-grams to available data fields and values, looking for both syntactic (e.g., misspellings) and semantic similarities (e.g., synonyms) employing Levenshtein distance~\cite{yujian2007normalized} and the Wu-Palmer similarity score~\cite{wu1994verb}, respectively.
The extracted items are subsequently used to support features like adding mark annotations (Figures~\ref{fig:scenario-1}B,~\ref{fig:scenario-2}B) and highlighting relevant portions of the chart while hovering over statements in the description (Figure~\ref{fig:scenario-1}A).

In addition to detecting data item references, the parser also classifies description statements into one of the five statement types.
We use a random forest classifier with BERT~\cite{sanh2019distilbert} to match a statement to one of the four semantic levels of text---\schemaSecondary{encoding}, \schemaSecondary{perceptual-trend}, \schemaSecondary{data-fact}, or \schemaSecondary{domain-specific}~\cite{lundgard2021accessible}.
If the classification probability for all four types is below 60\% \new{(an empirically set threshold)}, a statement is labeled as \schemaSecondary{other}.
The classifier is trained on a dataset of $2147$ chart description statements curated by Lundgard and Satyanarayan~\cite{lundgard2021accessible}.
Our choice for the classifier was based on comparing the results of 10-fold cross-validation between different techniques, including support vector machines~\cite{Cortes1995SupportVectorN}, random forests~\cite{breiman2001}, logistic regression~\cite{strother1967}, and na\"{i}ve Bayes~\cite{Duda1974PatternCA}.

\textbf{Chart.}
The parser also detects salient \schemaPrimary{DataItems} in the chart.
For instance, for bar charts, the parser shortlists up to three categories with the highest and lowest values.
For line charts, the system records time periods or specific timestamps with the most significant peaks and drops based on computing the smoothed z-scores, and so on.
These chart-specific heuristics to determine salient targets are derived from prior ``auto-insight'' generating visualization systems (e.g.,~\cite{cui2019datasite,wang2019datashot,srinivasan2018augmenting,demiralp2017foresight}) and research on mappings between analytic tasks and visualizations (e.g.,~\cite{amar2005low,schulz2013design,saket2018task}).
The salient items detected from the chart are subsequently used to suggest potential annotations and to generate text suggestions for verifying the description statements (e.g., Figure~\ref{fig:scenario-1}A).

\subsection{Recommendation Generation}
\label{sec:reco-generation}
\pluto~uses a combination of heuristics, text templates, and an LLM to suggest changes to the text and the chart.
The vignettes in \S\ref{sec:scenarios} illustrate the breadth of \pluto's recommendations, which can broadly be categorized into three groups: 1) \textit{full-text recommendations} to populate descriptions, titles, or text annotations, 2) \textit{description statement recommendations} to fine-tune or update an existing description, and 3) \textit{chart design recommendations} to ensure the chart is structurally aligned to its corresponding text.

\begin{figure*}[t!]
    \centering
    \includegraphics[width=.96\textwidth]{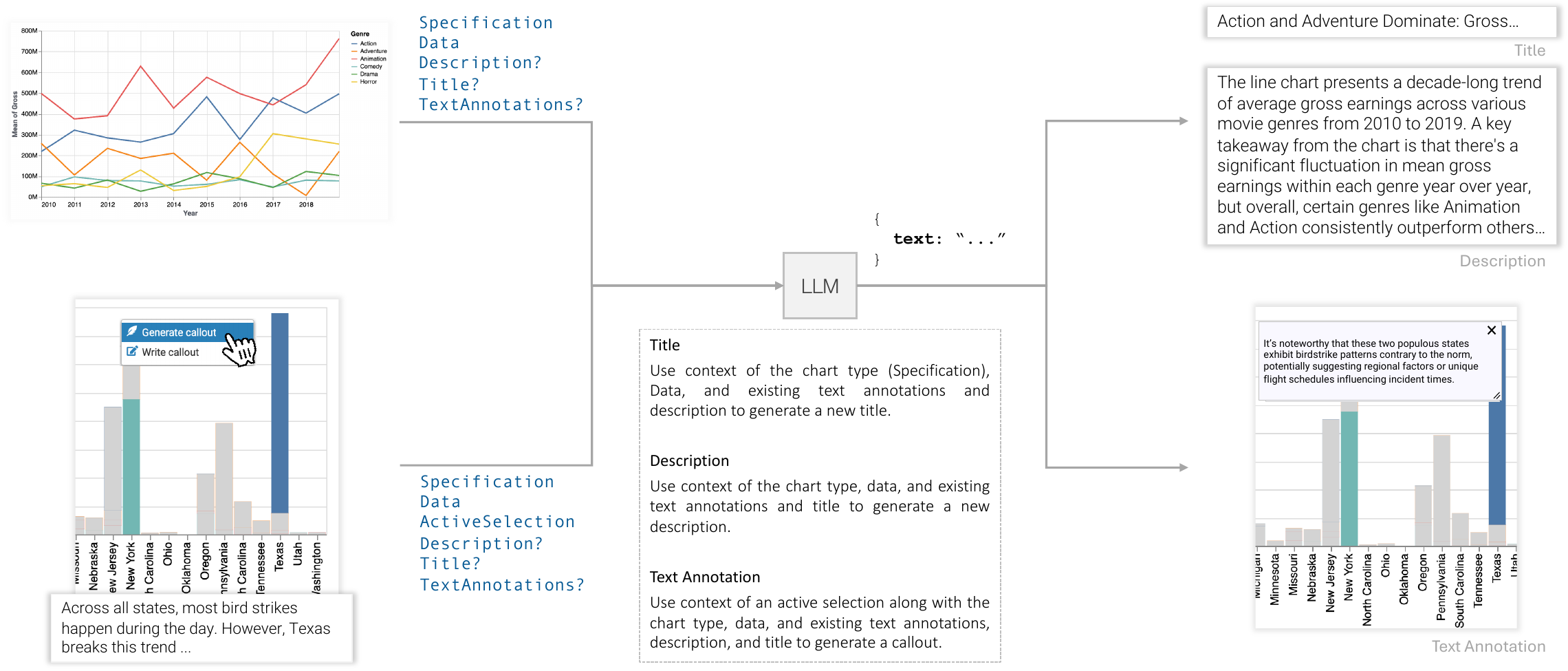}
    \caption{Overview of full-text recommendation generation. Given the context of the chart, data, and any existing text, \pluto~generates new text for the description, title, or annotations. \schemaPrimary{Input parameters} with a \schemaPrimary{?} are optional only used if available.}
    \Description[Overview of full-text recommendation generation.]{Given the context of the chart, data, and any existing text, Pluto uses an LLM to suggest the title, description, and text annotations.}
    \label{fig:full-text-recommendation}
\end{figure*}

\vspace{.5em}
\noindent{\large{\textbf{Full-text Recommendations}}
\vspace{.5em}

\noindent{}These recommendations are invoked using the {\small\faIcon{feather-alt}} \textbf{Generate} button and suggest text for the title, description, or text annotations (e.g., Figure~\ref{fig:teaser}A and Figure~\ref{fig:scenario-2}C).
All recommendations in this category are generated using the LLM, and Figure~\ref{fig:full-text-recommendation} presents an overview of the input/output for the recommendations.
The LLM prompts are provided as part of the supplementary material.

\textbf{Description.}
We use \schemaPrimary{StatementTypes} to systematically generate descriptions in \pluto.
Specifically, we provide the LLM with examples of the four statement types from Lundgard and Satyanarayan's dataset~\cite{lundgard2021accessible}.
Following the findings from Tang et al.'s qualitative analysis of the VisText chart caption dataset~\cite{tang2023vistext}, we prompt the LLM to generate a description with a constraint that the text should start with an \schemaSecondary{encoding} statement and is followed by at least one \schemaSecondary{preceptual-trend}.
This pattern follows the classic \textit{``Overview first''} mantra for visualization design~\cite{shneiderman2003eyes} and ensures the description talks about the chart and high-level takeaways before listing details of individual items and values.
An example of this constraint in play can be noticed in the generated description in Figure~\ref{fig:full-text-recommendation} where the first statement, ``\textit{This line chart...}'' describes the chart's encodings and the second highlights how ``\textit{...certain genres like Animation and Action consistently outperform others...}'' before talking about other lower-level observations from the chart.

By default, the LLM only uses the chart type and \schemaPrimary{Data} to generate a description.
However, if the chart contains an \schemaPrimary{ActiveSelection}, has \schemaPrimary{TextAnnotations}, or the user has entered a \schemaPrimary{Title}, these are also used as context for generating the description.

\textbf{Title.}
By default, the system uses a chart's \schemaPrimary{Specification} and \schemaPrimary{Data} to generate a title.
However, similar to generating descriptions, if there is additional context in the form of an \schemaPrimary{ActiveSelection}, \schemaPrimary{TextAnnotations}, or a \schemaPrimary{Description},\\ \pluto~leverages that information to generate a title that highlights the key message across the chart and previously added text.
For instance, the suggested title in Figure~\ref{fig:full-text-recommendation} contains \textit{Action} and \textit{Animation} since these are called out as focal entities in the \schemaPrimary{Description}.

\textbf{Text annotations.}
\pluto~also allows users to directly select items of interest on the chart and generate annotations based on the \schemaPrimary{ActiveSelection} (\textbf{DG2}).
An example of this type of text generation is shown in Figure~\ref{fig:full-text-recommendation}-bottom where the \schemaPrimary{TextAnnotation} is created based on the two selected bars for the states of \textit{New York} and \textit{Texas}, respectively.
The example also illustrates the effect of including previously entered text as context for the generation.
Specifically, notice that because the \schemaPrimary{Description} talks about Texas and New York deviating from the general trend, the generated annotation text also adopts that framing and phrasing (e.g., ``\textit{...contrary to the norm...}'') for consistency.
\newline

\noindent{}Since all the above recommendations leverage the chart type (inferred via the \schemaPrimary{Specification}) and \schemaPrimary{Data}, from an implementation standpoint, we pass the chart type and the data to the LLM only once in an initial context setting prompt when a chart is created.
Our choice to include the chart type as part of the context was motivated by our initial testing, during which we found that including the chart type improved the LLM's performance in terms of detecting the most relevant data patterns (e.g., trends for line charts, extremes for bar charts, correlation for scatterplots).

\vspace{.5em}
\noindent{\large\textbf{Description Statement Recommendations}}
\vspace{.5em}

\noindent{}Besides suggesting text from scratch, \pluto~also recommends adding or editing \schemaPrimary{Statements} within an existing \schemaPrimary{Description} (\textbf{DG3}).
Statement recommendations take different forms, including in-place suggestions to verify statement correctness (Figure~\ref{fig:scenario-1}A), statement addition/reordering recommendations presented to the side of an existing description (Figure~\ref{fig:interface}G), and in-place text completions (Figure~\ref{fig:teaser}B and Figure~\ref{fig:scenario-2}A).

\begin{figure*}[t!]
    \centering
    \includegraphics[width=\textwidth]{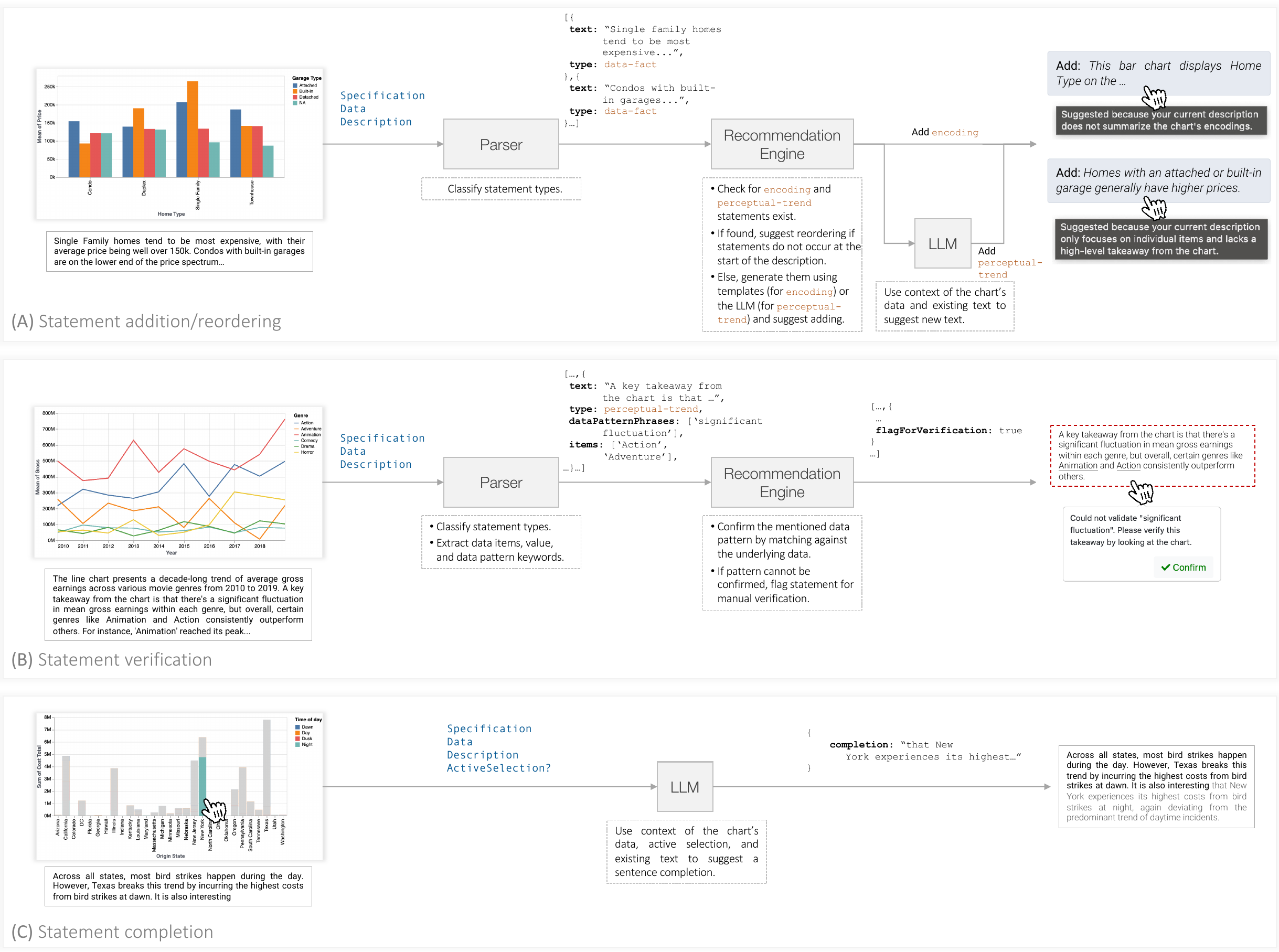}
    \caption{Overview of the description statement recommendations in \pluto. The system uses a combination of the chart's specification, data, the active description, and selections on the chart to recommend changes to the description.}
    \Description[Overview of description statement recommendations in Pluto.]{To generate statement addition/reordering suggestions, the system uses a combination of a text parser and a heuristic recommendation engine that inspects the statement types (e.g., encoding, perceptual-trend) to suggest content. A LLM is used when perceptual trend statements are suggested as part of the recommendations. For statement verification recommendations, the system uses a parser to identify statement items, extract data references, and subsequently checks these against the underlying data to flag a statement as needs verification or not. Lastly, for statement completion recommendations, the system passes the context of the current chart, description, and any active selections to the LLM to have it generate the statement text.}
    \label{fig:statement-recommendations}
\end{figure*}

\textbf{Statement addition and reordering.}
These recommendations are designed to ensure the description is semantically rich and has a good narrative structure.
Figure~\ref{fig:statement-recommendations} summarizes the logic used to generate statement recommendations.
Note that because the rules used to generate these recommendations are already baked into the description generation prompt described above, the statement addition/reordering recommendations typically appear only for manually entered descriptions.

To generate the recommendations, we follow the same guidelines applied to generate a description from scratch. \pluto~first checks for the presence of an \schemaSecondary{encoding} statement and at least one \schemaSecondary{perceptual-trend}.
If these are absent or placed after other statements, the system recommends adding or reordering these statements.
For instance, consider the example in Figure~\ref{fig:statement-recommendations}A.
Detecting that the input description lacks both \schemaSecondary{encoding} and \schemaSecondary{precentual-trend} statements, \pluto~suggests adding one statement of each type.
We initially also explored suggesting adding \schemaSecondary{data-facts}.
However, the formative studies and our testing revealed that these recommendations quickly became mundane and merely listed data values from the chart, leading to us subsequently disabling them.
Drawing on prior work~\cite{tang2023vistext}, we use a template-based approach to suggest \schemaSecondary{encoding} statements and invoke the LLM to suggest \schemaSecondary{perceptual-trends}.

To avoid overwriting the existing description, the recommendations are presented next to the description area instead of being directly applied to the text (Figure~\ref{fig:interface}G) (\textbf{DG5}).
Authors can preview the updated description with the suggested changes by hovering on the recommendations.
Furthermore, because these recommendations are heuristically generated, \pluto~also provides an explanation for why a recommendation was shown.
The output in Figure~\ref{fig:statement-recommendations}A shows examples of these explanations accompanying an \schemaSecondary{encoding} statement suggestion and a \schemaSecondary{preceptual-trend} statement suggestion generated when the input description only contains \schemaSecondary{data-facts}.

\textbf{Statement verification.}
Prior research has shown that both manually-written and LLM-generated descriptions can contain erroneous mentions of data trends or values~\cite{kim2023emphasischecker,tang2023vistext}. For instance, a category stated to have the highest value in the text may not actually be the category with the highest value in the chart.
Motivated by this prior work and the experts' feedback on earlier prototypes, we check for potentially incorrect data references in the text and flag them for authors to manually verify (\textbf{DG3}).

Figure~\ref{fig:statement-recommendations}B gives an overview of how \pluto~flags statements for verification.
Specifically, the system first checks if the statement contains one or more \schemaPrimary{DataItems} (typically found in \schemaSecondary{data-fact} and \schemaSecondary{perceptual-trend} statements) and the type of takeaway the statement calls out (e.g., min/max, trend, correlation).
\pluto~then uses this extracted information to validate the mentioned items and values against the underlying \schemaPrimary{Data}, flagging the statement for review if it fails to detect a match.
An example of this is shown in Figure~\ref{fig:statement-recommendations}B, where the sentence is flagged because the system is unable to confirm if the fluctuation in values is ``significant''.
Authors can click on a flagged statement to {\small\faIcon{check}} \textit{Confirm} its correctness.

\textbf{Statement text completion.}
In addition to retrospective recommendations on the description text, \pluto~also allows users to request text completion suggestions while writing their descriptions by pressing the \key{Tab} key.
To generate these completions, the system sends the current \schemaPrimary{Description} along with any \schemaPrimary{ActiveSelections} on the chart to the LLM and prompts it to complete or suggest the last sentence.
Examples of these suggestions can be seen in Figure~\ref{fig:scenario-2}A, where the system generates a text completion based on the \schemaPrimary{Description} alone and Figures~\ref{fig:teaser}B and~\ref{fig:statement-recommendations}C, where the completion is generated based on multimodal input, including the \schemaPrimary{description} text and the \schemaPrimary{ActiveSelection} of \textit{New York} on the chart.
If the text completion recommendation is invoked with an empty description, the system follows the same rules as it does when generating descriptions from scratch and starts by suggesting an \schemaSecondary{encoding} statement followed by a \schemaSecondary{perceptual-trend} before other statements.

\vspace{.5em}
\noindent{\large\textbf{Chart Design Recommendations}}
\vspace{.5em}

\begin{figure*}[t!]
    \centering
    \includegraphics[width=\textwidth]{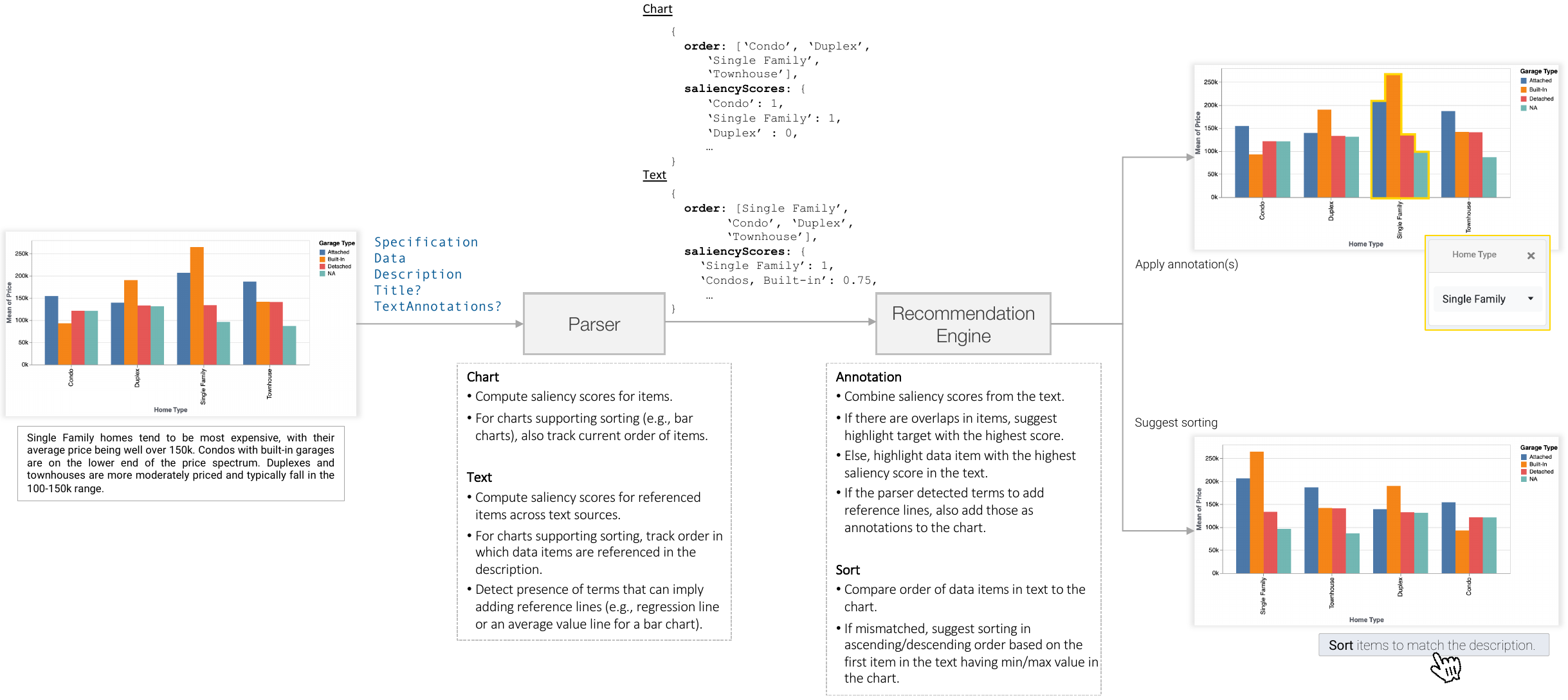}
    \caption{Summary of \pluto's process for recommending chart design changes based on the authored text. Given a chart and accompanying text, the system extracts data references from both the chart and the text, and compares the references to suggest potential design changes to make the chart more structurally aligned to the text.}
    \Description[Overview of the recommendation logic for suggesting chart design changes.]{To suggest design changes based on a given description, the system first inspects the chart to identify key data items (e.g., categories with highest values in bar charts). Comparing the data references in the input text to chart, the system suggests annotating the references in the chart, prioritizing items that have a higher salience in the chart. Additionally, the system also compares the order of the items in the text and the chart and if there is a difference, suggests sorting the chart to match the order of data references in the text.}
    \label{fig:chart-recommendations}
\end{figure*}

\noindent{}Following \textbf{DG1}, \pluto's recommendations are geared not only to improve the text but also to align the chart with the text, resulting in a better-combined reading experience.
Specifically, once a description is entered, the system generates two types of recommendations for updating the chart.

\textbf{Annotations.}
\pluto~recommends visual embellishments based on the description to help emphasize the key takeaways from the text in the chart following the approach summarized in Figure~\ref{fig:chart-recommendations}.
The recommended annotations are applied by default since they do not impact the chart's structure/layout, but authors are provided with controls to refine or remove the applied annotations (e.g., Figures~\ref{fig:interface}B and \ref{fig:scenario-1}B).

The system first extracts a list of potential \schemaPrimary{DataItems} from both the text and the chart and assigns a saliency score to these items based on saliency or ``interestingness'' metrics~\cite{demiralp2017foresight,wang2019datashot,srinivasan2018augmenting,lundgard2021accessible} (e.g., categories with extreme values in bar charts have higher saliency scores, time ranges in line charts with more variability have higher saliency than those with lower variability, targets mentioned in \schemaSecondary{perceptual-trend} statements are considered more salient than those referenced in \schemaSecondary{data-facts}).

Next, \pluto~checks for overlaps in \schemaPrimary{DataItems} extracted from the chart and the text to shortlist candidates for annotation based on a combined saliency score.
Checking for the combined saliency scores across the text and chart ensures that the emphasized items are important in both the underlying data and the author's interpretation of the chart conveyed via the text.
In cases where there are no overlaps between \schemaPrimary{DataItems} in the chart and the text, \pluto~defaults to adjusting the chart to match the author's description and adds an annotation for the most salient \schemaPrimary{DataItems} in the text.
An example of this is shown in Figure~\ref{fig:chart-recommendations} where the system highlights \textit{Home Type}$=$\textit{Single Family} since it has the highest average value (i.e., it is a salient data item in the chart) and is also explicitly called out in the description.

Besides annotating specific marks or regions on the chart, the system also adds overlay annotations based on references to aggregate values (e.g., adding a line to highlight the average value across categories in a bar chart or a regression line to emphasize the correlation between fields on a scatterplot).

\textbf{Sorting.}
During our initial testing and formative interviews, we noted that there was often a disparity in the order in which data items or marks appear on a chart and the order in which they are discussed in the text.
For example, consider the bar chart in Figure~\ref{fig:interface}C.
Although the bars are sorted alphabetically by home type, the description highlights takeaways starting with the home type having the highest value (i.e., \textit{Single Family} homes).
To enable a more aligned chart and text reading experience, \pluto~checks for such disparities and recommends sorting order changes for alignment.

Specifically, as summarized in Figure~\ref{fig:chart-recommendations}, the system extracts an ordered list of \schemaPrimary{DataItems} in the description and compares it to the items in the chart.
If the description starts by focusing on the highest or lowest category and the chart is not ordered to match that narrative, \pluto~suggests sorting the chart in a descending or ascending order, respectively.
An example of the sorting recommendation in action can be seen in Figure~\ref{fig:chart-recommendations} where applying the sort aligns the chart and text with both emphasizing \textit{Single Family} homes having the highest values and \textit{Condos} having the lowest values.

Unlike the annotation recommendations, however, sorting recommendations are presented next to a chart (Figure~\ref{fig:interface}F) and not applied by default to prevent an abrupt visual layout change by the system.
As shown in Figure~\ref{fig:chart-recommendations}, authors can preview the suggested order by hovering on the recommendation and clicking to apply that recommendation.
\section{Preliminary User Study}
\label{sec:study}

Using \pluto~as a design probe, we conducted a preliminary user study to assess the utility of the proposed interactive experience for authoring semantically aligned text and charts for data-driven communication.
Specifically, we focused on two goals: 1) understand if and how the system suggestions aid the joint authoring of text and charts, and 2) gather feedback on \pluto's current recommendations and features.

\subsection{Participants and Setup}

We recruited ten participants ($P1$-$P10$) through mailing lists at a data analytics software company.
\new{The recruitment call sought for individuals who use a combination of charts and text for data-driven communication.
Participation in the study was voluntary and} participants were recruited on a first-come, first-serve basis.
Seven participants rated themselves as visualization experts, and three participants had a moderate level of expertise in visualization. 
Regarding participants' professional backgrounds, six participants were solution architects, two participants worked as visualization consultants, and two were data analysts.
Of the ten participants, six participants reported that they used a combination of text and charts to communicate data once in a few weeks, two frequently used text and charts as part of their jobs, and two participants had only recently started using charts and text for communication as part of their new roles.
This mix of participant backgrounds w.r.t. chart and text authoring ensured that the study captured holistic feedback on \pluto~from varying users, including novices, moderate-level users, and experts.

Speaking about their existing authoring experience, all participants noted they manually inspected the chart and wrote a corresponding text blurb while sharing.
Five participants said they sometimes annotated a chart's screenshot with text during communication.
The two participants who frequently communicated using text and charts also noted using the dashboard authoring features in tools like Tableau.

All sessions were conducted remotely via the Zoom video conferencing software. The prototype was hosted on a local server on the experimenter's laptop. Participants were granted control over the experimenter's screen during the session, and all studies followed a think-aloud protocol. The audio, video, and on-screen actions were recorded for all sessions with permission from the participants.

\subsection{Procedure}

We initially considered an evaluation of \pluto~against an existing chart-and-text authoring tool. However, we did not find a freely available baseline that provided equivalent features to \pluto~in terms of supporting multimodal authoring, bidirectional editing of text and charts, and using the semantic structure of descriptions~\cite{lundgard2021accessible} during text recommendation.
We also considered an ablation study to compare the system's output to that from the underlying GPT-4 model.
However, we did not see value in this approach as \pluto's utility stems from the integration of multimodal interactions and recommendations, and not one standalone feature focused on text generation.

We ultimately decided on a qualitative study where all participants interact with \pluto~and perform the same set of tasks as this would allow us to observe usage patterns and assess the utility of the recommendations.
Sessions lasted between 44-57 minutes ($\mu$: 53 min., \new{$\sigma$: 4 min.}) and were organized as follows:

\vspace{.5em}
\noindent\textbf{Introduction} [$\sim$10min]:
Participants were given an overview of the study and asked to share their background information. This briefing was followed by an introduction to \pluto's interface and key features. We used a simple bar chart about US college costs as a running example for the introduction.

\vspace{.5em}
\noindent\textbf{Tasks} [$\sim$30min]:
Participants were presented with three charts and were asked to write complementary text (including title, description, and annotations) for the chart so they could share their findings with others who may be interested in the data.
To ensure the tasks were realistic and of appropriate difficulty, we consulted the two experts from the formative study (Section~\ref{sec:design-goals}) and conducted three pilot studies with participants from academic and journalism backgrounds.

The three charts included a multi-series line chart about movie earnings by genre, a stacked bar chart about costs incurred by airplane bird strikes in the US, and a grouped bar chart about US housing prices shown in Figures~\ref{fig:teaser}A, \ref{fig:scenario-2}A, and \ref{fig:interface}D, respectively.
\new{We selected these chart types based on their frequency of use in prior chart-and-text authoring systems (e.g.,~\cite{sultanum2023datatales,chen2022crossdata,choi2022intentable,kim2023emphasischecker,lin2023inksight}) as well as the general prevalence of bar and line charts across visualization platforms (e.g.,~\cite{battle2018beagle,purich2023toward}) that are commonly leveraged for data-driven communication.}
The order of charts was randomized across participants. 
We asked the participants to use the system as they saw fit (e.g., start with auto-generated text, manually write text and use the suggestions to edit, or manually compose the chart and text without using system recommendations).

\vspace{.5em}
\noindent\textbf{Debrief} [$\sim$10min]:
Sessions concluded with a semi-structured interview discussing the overall experience, support for different authoring tasks, and areas for improvement. Participants also filled out a questionnaire rating the quality and utility \pluto's recommendations and features.

\subsection{Results}

\begin{figure}[t!]
    \centering
    \includegraphics[width=\linewidth]{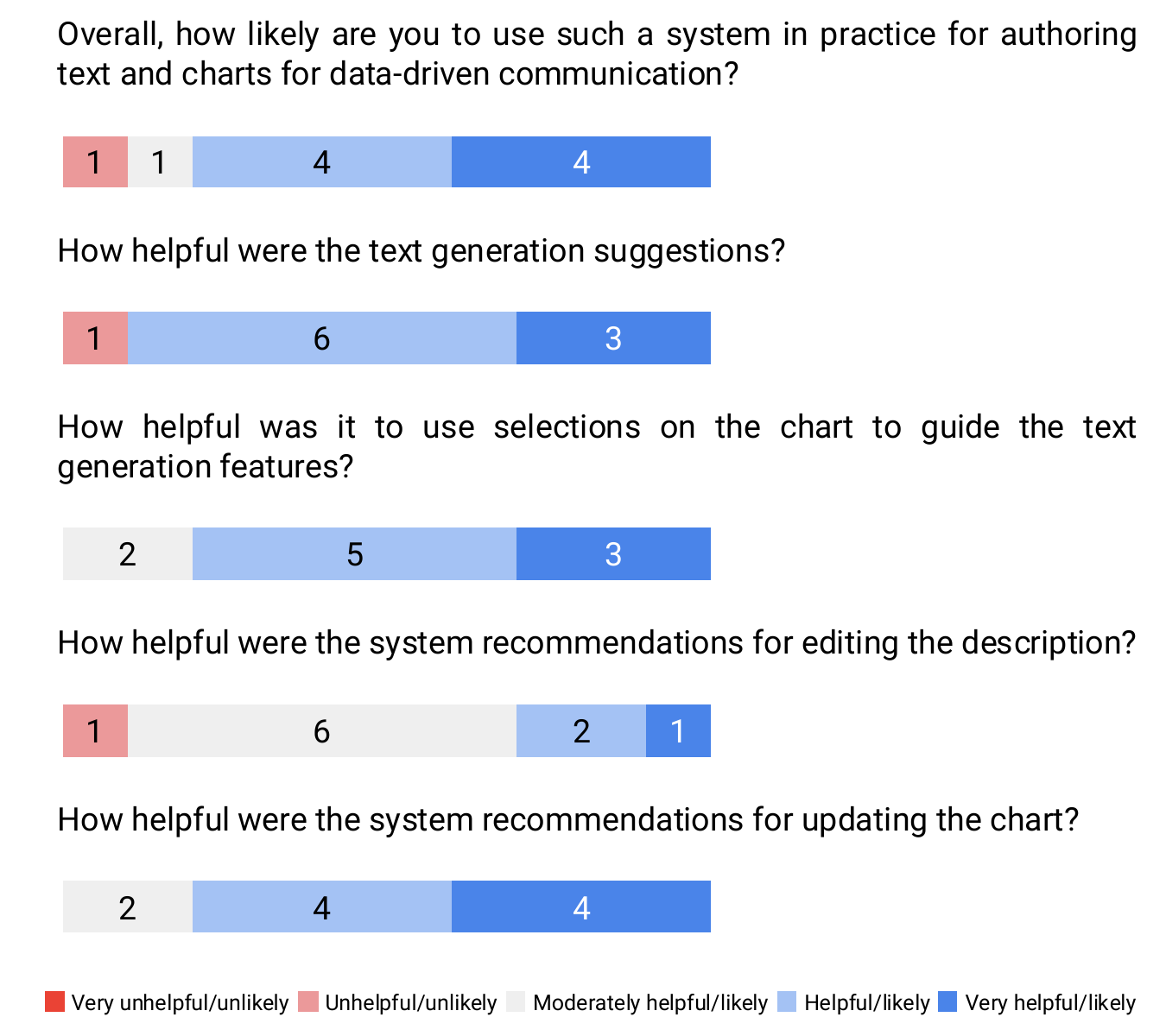}
    \caption{Participant responses to post-session questions about \pluto's recommendations.}
    \Description[Participant responses to post-session questionnaire]{Overall, participants were generally positive about Pluto's recommendations, particularly commending the use of selection to guide suggestions and the recommendations for updating the chart design based on the text. Detailed responses are as follows. Question: Overall, how likely are you to use such a system in practice for authoring text and charts for data-driven communication? Responses: 4 participants said "very likely", 4 said "likely", 1 said "moderately likely" and 1 said "unlikely".
    Question: How helpful were the text generation suggestions? Responses: 3 participants said "very helpful", 6 said "helpful", and 1 said "unhelpful". Question: How helpful was it to use selections on the chart to guide the text generation features? Responses: 3 participants said "very helpful", 5 said "helpful", and 2 said "moderately helpful". Question: How helpful were the system recommendations for editing the description? Responses: 1 participant said "very helpful", 2 said "helpful", 6 said "moderately helpful", and 1 said "unhelpful". Question: How helpful were the system recommendations for updating the chart? Responses: 4 participants said "very helpful", 4 said "helpful", and 2 said "moderately helpful".}
    \label{fig:responses}
\end{figure}

Overall, participants were receptive to \pluto's features and recommendations, noting they would use such a system in practice for data-driven communication (Figure~\ref{fig:responses}). We summarize general themes from participant feedback.

\vspace{.5em}
\noindent\textbf{Full-text recommendations are helpful for bootstrapping.}
Participants generally found the text generation recommendations useful, with nine participants rating them as `helpful' or `very helpful' (Figure~\ref{fig:responses}).
Participants noted that these suggestions were particularly useful for bootstrapping the authoring process. For instance, commenting on the generated description, $P1$ said, ``\textit{I would love to use these as a first draft. Just helps get my juices flowing.}''
However, participants' feedback on the quality of generated descriptions was mixed, with some participants (P2, P4, P7) finding the suggested text too verbose.
Both $P4$ and $P7$, for instance, noted that they would prefer the system generate a bullet list of key ``insights," allowing them to use the insights to manually craft a narrative.
$P2$ and $P4$ (both frequent users of text and charts for communication) also stated they would like more control over the generated text in terms of its verbosity and writing style (e.g., configuring the text to have a more ``casual'' versus ``formal'' tone depending on the communication context).

Feedback on the suggested titles was unanimously positive, however.
For example, commending a suggested title for the house pricing chart, $P8$ said, ``\textit{Terms like `Influence of' really makes chart feel less templated and more informative.}''
Upon seeing the title ``Action and Adventure Dominate:...", $P5$ noted, ``\textit{I am very impressed with the title. It's almost like it took all my changes and gave me a summary."}
Participants were also pleasantly surprised by the quality of text completions for individual sentences (e.g., Figure~\ref{fig:scenario-2}A) stating ``\textit{the text completion followed my lead very well}" ($P6$).

\vspace{.5em}
\noindent\textbf{Using selections to guide recommendations saves time and instills confidence.}
Overall, participants appreciated the ability to directly select items on the chart to drive text recommendations, with 8/10 participants noting this feature was `helpful' or `very helpful' (Figure~\ref{fig:responses}).
The positive feedback for the multimodal text generation feature often stemmed from its ability to assist in faster writing and to adjust the scope of the generated text.

$P8$, for instance, particularly appreciated the ability to use chart selections to drive auto-complete and annotations (e.g., Figures~\ref{fig:teaser}B and \ref{fig:scenario-2}C) and said, ``\textit{It was nice to be able to point at things and have the system give the text. It saved me a lot of time.}"
P2 and P4, who found the description generation feature too verbose, switched to using selection-based text generation, with P4 stating that ``\textit{the selection at least ensured the generated text was about something I want to talk about.}''

\vspace{.5em}
\noindent\textbf{Description editing recommendations are primarily useful for validation.}
Participants' reactions to recommendations for verifying and editing an entered description (e.g., Figures~\ref{fig:scenario-1}A and \ref{fig:interface}G) were more neutral (Figure~\ref{fig:responses}, Q4).

Seven participants explicitly noted that they appreciated that the system flagged text that needed verification.
P6, for instance, said, ``\textit{Humans are lazy. Having that extra step is going to save a lot of people a lot of embarrassment.}''
Noting that the verification suggestions made him more critical, $P3$ said, ``\textit{It's important that we think about what charts are saying...and it's referring back and making sure. It's definitely a step in the right direction.}''
The remaining participants either felt the suggestions should not appear for manually written text ($P4$) or that statements could be flagged more conservatively, reducing the work on the authors' part ($P7$, $P8$).

Participants used recommendations to add or reorder text in the description only three times across all sessions. Participants commented that these recommendations were too ``obvious'' (particularly referring to the \schemaSecondary{encoding} statement suggestions).
$P9$, for instance, ``\textit{This [\schemaSecondary{encoding} statement] suggestion is too generic and not as interesting as the others that tell me key points about the data.}''
However, all participants noted that the recommendations were appropriately placed on the side and did not impede their workflow (\textbf{DG5}).

\vspace{.5em}
\noindent\textbf{Chart annotation and design recommendations foster an integrated reading experience.}
Participants were generally very impressed by \pluto's suggestions to annotate or sort the chart based on an entered description, with 8/10 participants rating this feature as `helpful' or `very helpful.'

For example, even $P7$, who was generally critical about the other features, exclaimed, ``\textit{Loved that just loved that! It's easy to forget the chart when writing because I know what I should be focusing on, but someone else looking at the chart may not.}''
Appreciating the ability to adjust the system-suggested annotations further (\textbf{DG5}), P1 commented, ``\textit{It's great that it highlighted some elements in the chart based on my text. It made me see things from a reader's perspective and go back and make additional changes.}''

\vspace{.5em}
\noindent\textbf{Usage patterns.}
As typical with mixed-initiative interfaces, there was a constant back and forth between the participants' authoring actions and the system recommendations. We observed four high-level usage strategies around how participants started the authoring process by writing descriptions. We summarize these strategies below, as they can help inform the user experience of future systems\footnote{Note that some participants adopted different strategies across tasks, resulting in the participant count across strategies adding up to more than 10}.

\begin{tightItemize}
    \item \textit{Generate then edit.}
    The most common strategy across participants (7/10) was to start with auto-generated descriptions ({\small\faIcon{feather-alt}} \textbf{Generate}).
    Once the description text was generated, participants would either first peruse through it and make edits or directly request suggestions for improvements to the text and the chart ({\small{\faIcon[regular]{lightbulb}}} \textbf{Suggest}).
    
    \vspace{.5em}
    \item \textit{Guide text completion.}
    Four participants started writing their descriptions leveraging the in-place text competition suggestions (via the \key{Tab} key).
    Participants would typically mention a data entity (e.g., New York) or a narrative hook (e.g., ``\textit{however},'' ``\textit{but},'' ``\textit{sadly}'') in their text and have the system suggest the remaining statement that they would use as-is or edit further.
    The {\small{\faIcon[regular]{lightbulb}}} \textbf{Suggest} feature was primarily used to verify the statements and get chart annotation recommendations to complement the text.

    \vspace{.5em}
    \item \textit{Clipboard text generation.}
    On two occasions, participants used the system text suggestions to create a clipboard of ideas.
    Specifically, participants started by selecting visually salient entities on the chart and asking \pluto~to generate a series of text annotations. Subsequently, the participants went through these annotations and either edited and kept them on the chart, moved them to the description, or deleted them.
    The {\small{\faIcon[regular]{lightbulb}}} \textbf{Suggest} feature was used after this initial drafting of the description and text annotations to get further editing suggestions for the chart.

    \vspace{.5em}
    \item \textit{Manual writing with chart design recommendations.}
    Two participants (both experts at communicating data using text and charts) typically started their process by manually drafting the description and then asking the system to {\small{\faIcon[regular]{lightbulb}}} \textbf{Suggest} improvements. These participants also noted that they utilized the suggest feature to obtain chart annotations and design suggestions based on their input text (\textbf{DG1}).
\end{tightItemize}
\section{Discussion}

\begin{figure*}[t!]
    \centering    \includegraphics[width=.6\linewidth]{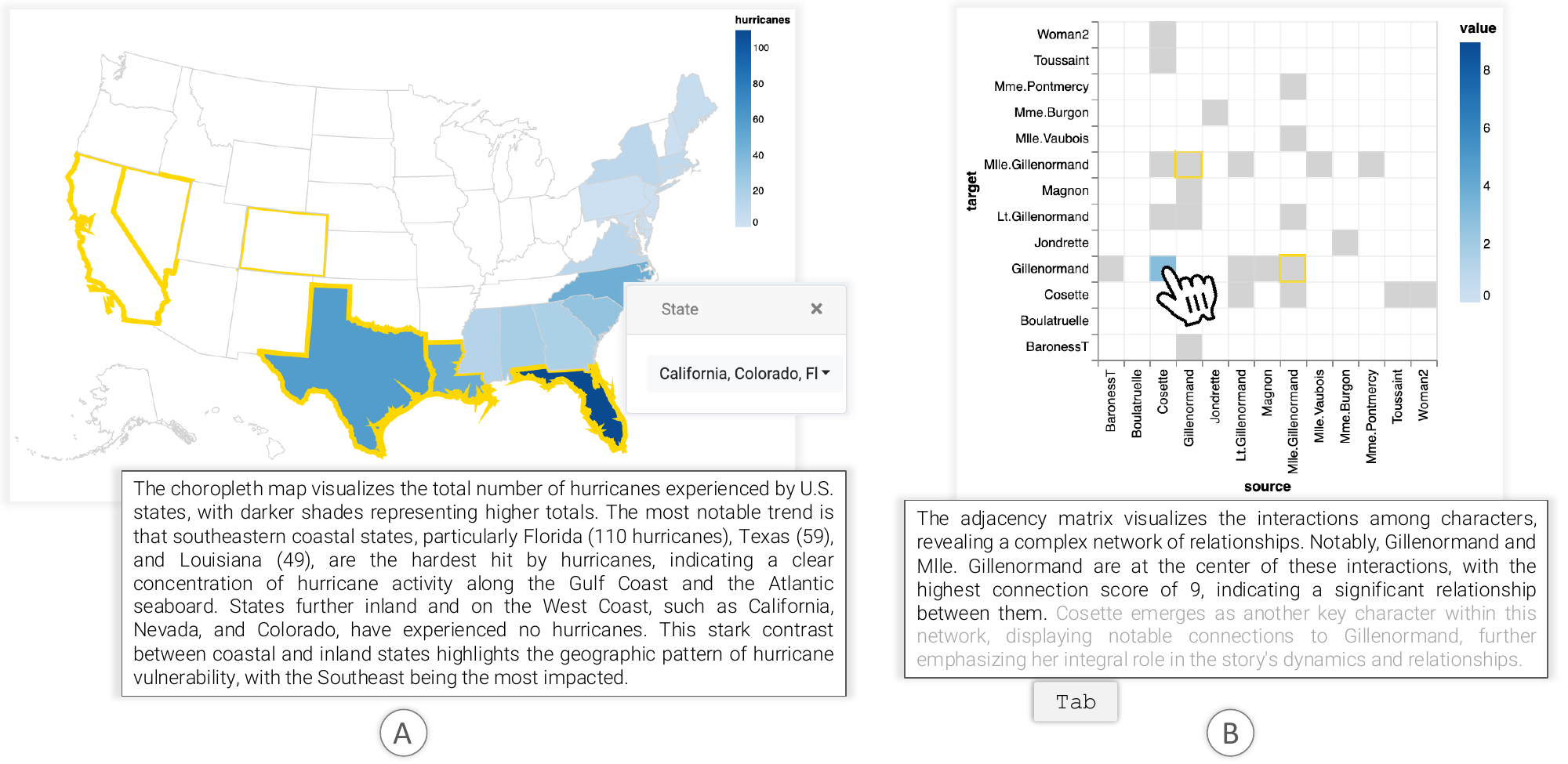}
    \caption{Examples illustrating the extensibility afforded by the proposed conceptual schema. Here, \pluto's text generation and recommendation modules are used \emph{as-is} to author text for a geographic map and an adjacency matrix. In (A), the system generates a description based on the map and subsequently highlights six states in the chart that are emphasized in the text. In (B), \pluto~suggests a sentence completion based on multimodal input of the previously drafted description and a user selection on the chart.}
    \Description[Two examples showing how the underlying schema used to develop Pluto can be extended to other charts.]{The first example illustrates the system is able to generate a textual description based on a choropleth map. The example also shows that some states are highlighted in the map because they are referenced in the generated text. The second example shows a sentence completion recommendation generated in response to clicking on a cell in a heatmap, further illustrating the system's ability to support charts that were beyond the initial set of visualizations implemented in the tool.}
    \label{fig:extensibility-examples}
\end{figure*}

\subsection{Grammar-based Approaches for Chart \& Text Interfaces}

While we currently implement only a set of charts and recommendations in \pluto, the underlying schema (Figure~\ref{fig:schema}) is not limited to this set.
Since the framework is based on the underpinning data and conceptual elements of the chart and text, the presented schema can be generalized to other cases.
Figure~\ref{fig:extensibility-examples} highlights examples of this generalization by illustrating \pluto's~text and embellishment suggestions for a geographic map and a graph dataset represented as a network matrix.

Furthermore, the idea of breaking individual \schemaPrimary{Statements} into \schemaPrimary{Text}, \schemaPrimary{StatementType}, and \schemaPrimary{DataItems} enables interactive text-chart linking (Figure~\ref{fig:scenario-1}A) as well as targeted recommendations for description statements (Figure~\ref{fig:statement-recommendations}).
Beyond these recommendations, however, this breakdown can also be leveraged more generally to design better linting systems for writing text for charts.
For example, by inspecting the \schemaPrimary{Text} and \schemaPrimary{StatementType}s in a given description and applying findings from studies on accessible visualization descriptions~\cite{yaneva2015accessible,lundgard2021accessible,jung2021communicating,kim2023explain}, visualization systems can flag inaccessible descriptions and suggest improvements to make the descriptions more accessible.

While these are just examples, leveraging logical concepts such as those in the presented schema (Figure~\ref{fig:schema}) affords a compelling opportunity to create a unifying grammar to capture the multimodal nature of chart + text authoring interfaces.
Besides streamlining the interface and interaction design of future tools, such a grammar that is centered around abstract concepts underpinning text and charts can also facilitate effective communication with LLMs by allowing systems to only share required task-specific context with the models in a structured representation.

\subsection{Design Considerations}

Based on the study observations and feedback, we derive four design considerations for future systems exploring AI-based suggestions to assist unified text and chart authoring.

\vspace{.5em}
\noindent\textbf{Format and phrasing of text are as important as its content.}
Participants recognized the value in text generation and noted that the generated text often picked up on the salient data points and trends in the data. However, the two participants who frequently used text and charts for communication, in particular, critiqued the verbosity and tone of the generated text, indicating that the text was ``\textit{too formal or complex}'' for their consumers. Future systems should explore providing users with control over the configuration of the properties of the generated text (e.g., choosing between paragraphs and bullets, adjusting the level of verbosity, and setting the tone or writing style)~\cite{louis2014}.

\vspace{.5em}
\noindent\textbf{Leverage multimodal context during text generation.}
Participants particularly appreciated text suggestions that were based on chart selections or built upon previously authored text.
$P5$, for instance, referred to the interaction experience of having some text and then using the chart in tandem to guide the system (Figure~\ref{fig:teaser}B) as being a ``\textit{smooth and controlled authoring flow.}'' To support similar fluid and coherent authoring experiences, future systems should consider multimodal context from both the chart and previously written text when suggesting new text.

\vspace{.5em}
\noindent\textbf{Include techniques to help verify the text with the chart.}
Participants appreciated the interactive highlighting and statement verification features in \pluto~(Figure~\ref{fig:scenario-1}A), noting that the features encouraged them be more critical of the text (regardless of whether it was written by them or was system-generated).
With the growing prevalence of generated text, future systems should continue incorporating such interactive verification features to mitigate false statements and enhance synergy between the text and the chart.

\vspace{.5em}
\noindent\textbf{Adjust chart design to align with the text description.}
Unlike prior systems that focus on the unidirectional task of generating text from charts, \pluto~introduces a bidirectional flow by also recommending changes to the chart design based on the text (Figure~\ref{fig:chart-recommendations}).
We observed that participants extensively used and appreciated these suggestions, commenting that the chart design recommendations helped them gain a better reader's perspective.
To this end, future systems should continue to explore techniques to leverage the bidirectional flow of information between text and charts
and generate chart design suggestions for an integrated reading experience.

\section{Limitations and Future Work}
\noindent\textbf{Incorporate data context into generated text.}
\pluto~currently only considers the active chart and data fields while suggesting text.
However, since the recommendations are provided within a general-purpose visualization specification tool, the system can access other data fields not displayed in the active chart (Figure~\ref{fig:interface}A).
For instance, during the study, viewing the generated text for the bird strikes chart (Figure~\ref{fig:scenario-2}A), $P8$ said, ``\textit{I wish it could generate some text explaining why the costs were high based on the number of incidents.}''
Although $P8$ was able to work around this limitation by creating the second chart, inspecting the new visualization, and returning to the original chart, automatically considering other fields as part of the generated text is an open area for future work.
Such dataset-level text (as opposed to chart-level text) can also help make the authored descriptions semantically rich by including more \schemaSecondary{domain-specific} statements.

\vspace{.5em}
\noindent\textbf{Text formatting recommendations.}
The current implementation leans heavily on content generation; however, future iterations should also focus on providing sophisticated formatting suggestions that can add expressivity to the text based on the semantic levels.
For example, titles can be rendered more prominently with annotations and footnotes shown to support the chart. Complementing chart design suggestions and providing structural suggestions for the description, such as including line breaks or using bullet points instead of paragraphs, can also help improve readability.

\vspace{.5em}
\noindent\textbf{Supporting multiple views and articles.} \pluto~currently supports authoring a single chart and assumes the resulting chart and text are static.
While such an approach covers a popular scenario for data-driven communication as evidenced by prior research focusing on this setup (e.g.,~\cite{choi2022intentable,obeid2020chart,liu2020autocaption,tang2023vistext,hsu2021scicap,stokes2022striking}), people also use dashboards and interactive storytelling articles when communicating with charts~\cite{sarikaya2018we,segel2010narrative}. Investigating support for multiple views and operationalizing \pluto's recommendations in more expressive interactive data-driven article authoring tools such as Idyll Studio~\cite{conlen2021idyll} is an open topic for future work.

\vspace{.5em}
\noindent\textbf{Manage data sharing and combine heuristics with LLMs.}
As discussed in \S\ref{sec:reco-generation}, we currently pprovide the context of the chart's specification and the data to the LLM.
However, depending on the size of the data and or the usage context,  sharing data directly with the LLM may not always be feasible.
Exploring alternatives to generate text without sharing the underlying data is an important direction for future work.
For instance, one possible approach could be to use heuristics-based approaches and prior knowledge of analytic tasks (e.g.,~\cite{manelski-krulee-1965-heuristic,snowy2021,kim2023emphasischecker}) to identify key trends from a chart and then leverage the LLM for appropriately consolidating the extracted information into a coherent text narrative.

\vspace{.5em}
\noindent\textbf{Conduct longitudinal evaluation across data domains.}
The combination of two expert interviews, three pilots, and an evaluation study with ten participants helped us validate \pluto~as a proof-of-concept for the general idea of a mixed-initiative interface for authoring semantically aligned charts and text.
\new{However, this evaluation is only preliminary, and assessing the practical value of a tool like \pluto~deems more longitudinal studies spanning different data domains and individuals with varying levels of expertise~\cite{lam2012,shneiderman2006strategies}.}

\vspace{.5em}
\noindent\new{\textbf{Incorporating safeguards for machine failures.}
With the current implementation of \pluto, we utilize a pretrained LLM to generate text and a heuristic parser to analyze the generated text for additional suggestions.
To make users aware of potential errors in either of these steps, we provide interactive visual previews that can be viewed before accepting system suggestions (e.g., Figure~\ref{fig:scenario-1}A and Figure~\ref{fig:chart-recommendations}).
While \pluto~serves as a proof-of-concept, developing a more generalized system requires a deeper investigation of the types of errors that might be generated both at the LLM and the parser level.
Categorizing and understanding the distribution of such errors can ultimately help develop better recommendation modules, but also design interface and interaction strategies to mitigate system errors.
}
\section{Conclusion}

The integration of AI-driven recommendations with user-driven customization presents a promising direction toward authoring more semantically aligned charts and text. This work introduces \pluto, a tool that incorporates a semantic and interactive framework that is informed by a nuanced understanding of the interplay between text and visuals. The system's recommendations, ranging from text generation to annotation and title suggestions, demonstrate the viability of a mixed-initiative approach to support richer chart-and-text authoring. 
A preliminary evaluation of the system revealed several key takeaways: participants highly valued the ability to use multimodal input for generating text, found text generation recommendations particularly useful for initiating the authoring process, and appreciated the system's ability to suggest annotations and chart design modifications based on the entered description. These findings underscore the potential of systems like \pluto~to facilitate the creation of more semantically coherent and integrated text and chart content. We envision a continued exploration of this research space, extending \pluto's capabilities to accommodate a broader array of visualization types and narrative styles, wherein text and visuals are not mere complements but coequals in data-driven communication.
\begin{acks}
\new{This work was supported, in part, by the National Science Foundation under III-1900991.}
\end{acks}

\bibliographystyle{ACM-Reference-Format}
\bibliography{references}


\begin{thebibliography}{66}


\ifx \showCODEN    \undefined \def \showCODEN     #1{\unskip}     \fi
\ifx \showDOI      \undefined \def \showDOI       #1{#1}\fi
\ifx \showISBNx    \undefined \def \showISBNx     #1{\unskip}     \fi
\ifx \showISBNxiii \undefined \def \showISBNxiii  #1{\unskip}     \fi
\ifx \showISSN     \undefined \def \showISSN      #1{\unskip}     \fi
\ifx \showLCCN     \undefined \def \showLCCN      #1{\unskip}     \fi
\ifx \shownote     \undefined \def \shownote      #1{#1}          \fi
\ifx \showarticletitle \undefined \def \showarticletitle #1{#1}   \fi
\ifx \showURL      \undefined \def \showURL       {\relax}        \fi
\providecommand\bibfield[2]{#2}
\providecommand\bibinfo[2]{#2}
\providecommand\natexlab[1]{#1}
\providecommand\showeprint[2][]{arXiv:#2}

\bibitem[veg(2014)]%
        {vega2014visualization}
 \bibinfo{year}{2014}\natexlab{}.
\newblock \bibinfo{title}{{Vega: A Visualization Grammar}}.
\newblock \bibinfo{howpublished}{\url{https://vega.github.io/vega/}}.
\newblock
\newblock
\shownote{Accessed: January 2025}.


\bibitem[Achiam et~al\mbox{.}(2023)]%
        {achiam2023gpt}
\bibfield{author}{\bibinfo{person}{Josh Achiam}, \bibinfo{person}{Steven Adler}, \bibinfo{person}{Sandhini Agarwal}, \bibinfo{person}{Lama Ahmad}, \bibinfo{person}{Ilge Akkaya}, \bibinfo{person}{Florencia~Leoni Aleman}, \bibinfo{person}{Diogo Almeida}, \bibinfo{person}{Janko Altenschmidt}, \bibinfo{person}{Sam Altman}, \bibinfo{person}{Shyamal Anadkat}, {et~al\mbox{.}}} \bibinfo{year}{2023}\natexlab{}.
\newblock \showarticletitle{{GPT-4 Technical Report}}.
\newblock \bibinfo{journal}{\emph{arXiv preprint arXiv:2303.08774}} (\bibinfo{year}{2023}).
\newblock


\bibitem[Alam et~al\mbox{.}(2023)]%
        {alam2023seechart}
\bibfield{author}{\bibinfo{person}{Md~Zubair~Ibne Alam}, \bibinfo{person}{Shehnaz Islam}, {and} \bibinfo{person}{Enamul Hoque}.} \bibinfo{year}{2023}\natexlab{}.
\newblock \showarticletitle{{SeeChart: Enabling Accessible Visualizations Through Interactive Natural Language Interface for People with Visual Impairments}}. In \bibinfo{booktitle}{\emph{Proceedings of the 28th International Conference on Intelligent User Interfaces}}. \bibinfo{pages}{46--64}.
\newblock


\bibitem[Amar et~al\mbox{.}(2005)]%
        {amar2005low}
\bibfield{author}{\bibinfo{person}{Robert Amar}, \bibinfo{person}{James Eagan}, {and} \bibinfo{person}{John Stasko}.} \bibinfo{year}{2005}\natexlab{}.
\newblock \showarticletitle{{Low-Level Components of Analytic Activity in Information Visualization}}. In \bibinfo{booktitle}{\emph{IEEE Symposium on Information Visualization, 2005. INFOVIS 2005.}} IEEE, \bibinfo{pages}{111--117}.
\newblock


\bibitem[Basole and Major(2024)]%
        {basole2024generative}
\bibfield{author}{\bibinfo{person}{Rahul~C Basole} {and} \bibinfo{person}{Timothy Major}.} \bibinfo{year}{2024}\natexlab{}.
\newblock \showarticletitle{{Generative AI for Visualization: Opportunities and Challenges}}.
\newblock \bibinfo{journal}{\emph{IEEE Computer Graphics and Applications}} \bibinfo{volume}{44}, \bibinfo{number}{2} (\bibinfo{year}{2024}), \bibinfo{pages}{55--64}.
\newblock


\bibitem[Battle et~al\mbox{.}(2018)]%
        {battle2018beagle}
\bibfield{author}{\bibinfo{person}{Leilani Battle}, \bibinfo{person}{Peitong Duan}, \bibinfo{person}{Zachery Miranda}, \bibinfo{person}{Dana Mukusheva}, \bibinfo{person}{Remco Chang}, {and} \bibinfo{person}{Michael Stonebraker}.} \bibinfo{year}{2018}\natexlab{}.
\newblock \showarticletitle{\new{Beagle: Automated Extraction and Interpretation of Visualizations from the Web}}. In \bibinfo{booktitle}{\emph{Proceedings of the 2018 CHI Conference on Human Factors in Computing Systems}}. \bibinfo{pages}{1--8}.
\newblock


\bibitem[Breiman(2001)]%
        {breiman2001}
\bibfield{author}{\bibinfo{person}{Leo Breiman}.} \bibinfo{year}{2001}\natexlab{}.
\newblock \showarticletitle{{Random Forests}}.
\newblock \bibinfo{journal}{\emph{Machine Learning}} \bibinfo{volume}{45}, \bibinfo{number}{1} (\bibinfo{date}{Oct} \bibinfo{year}{2001}), \bibinfo{pages}{5–32}.
\newblock
\showISSN{0885-6125}
\urldef\tempurl%
\url{https://doi.org/10.1023/A:1010933404324}
\showDOI{\tempurl}


\bibitem[Chen and Xia(2022)]%
        {chen2022crossdata}
\bibfield{author}{\bibinfo{person}{Zhutian Chen} {and} \bibinfo{person}{Haijun Xia}.} \bibinfo{year}{2022}\natexlab{}.
\newblock \showarticletitle{\new{Crossdata: Leveraging Text-Data Connections for Authoring Data Documents}}. In \bibinfo{booktitle}{\emph{Proceedings of the 2022 CHI Conference on Human Factors in Computing Systems}}. \bibinfo{pages}{1--15}.
\newblock


\bibitem[Choi and Jo(2022)]%
        {choi2022intentable}
\bibfield{author}{\bibinfo{person}{Jiwon Choi} {and} \bibinfo{person}{Jaemin Jo}.} \bibinfo{year}{2022}\natexlab{}.
\newblock \showarticletitle{{Intentable: A Mixed-initiative System for Intent-based Chart Captioning}}. In \bibinfo{booktitle}{\emph{2022 IEEE Visualization and Visual Analytics (VIS)}}. IEEE, \bibinfo{pages}{40--44}.
\newblock


\bibitem[Conlen et~al\mbox{.}(2021)]%
        {conlen2021idyll}
\bibfield{author}{\bibinfo{person}{Matthew Conlen}, \bibinfo{person}{Megan Vo}, \bibinfo{person}{Alan Tan}, {and} \bibinfo{person}{Jeffrey Heer}.} \bibinfo{year}{2021}\natexlab{}.
\newblock \showarticletitle{{Idyll Studio: A Structured Editor for Authoring Interactive \& Data-Driven Articles}}. In \bibinfo{booktitle}{\emph{The 34th Annual ACM Symposium on User Interface Software and Technology}}. \bibinfo{pages}{1--12}.
\newblock


\bibitem[Cortes and Vapnik(1995)]%
        {Cortes1995SupportVectorN}
\bibfield{author}{\bibinfo{person}{Corinna Cortes} {and} \bibinfo{person}{Vladimir~Naumovich Vapnik}.} \bibinfo{year}{1995}\natexlab{}.
\newblock \showarticletitle{{Support-Vector Networks}}.
\newblock \bibinfo{journal}{\emph{Machine Learning}}  \bibinfo{volume}{20} (\bibinfo{year}{1995}), \bibinfo{pages}{273--297}.
\newblock
\urldef\tempurl%
\url{https://api.semanticscholar.org/CorpusID:52874011}
\showURL{%
\tempurl}


\bibitem[Cui et~al\mbox{.}(2019)]%
        {cui2019datasite}
\bibfield{author}{\bibinfo{person}{Zhe Cui}, \bibinfo{person}{Sriram~Karthik Badam}, \bibinfo{person}{M~Adil Yal{\c{c}}in}, {and} \bibinfo{person}{Niklas Elmqvist}.} \bibinfo{year}{2019}\natexlab{}.
\newblock \showarticletitle{{Datasite: Proactive Visual Data Exploration with Computation of Insight-based Recommendations}}.
\newblock \bibinfo{journal}{\emph{Information Visualization}} \bibinfo{volume}{18}, \bibinfo{number}{2} (\bibinfo{year}{2019}), \bibinfo{pages}{251--267}.
\newblock


\bibitem[Demiralp et~al\mbox{.}(2017)]%
        {demiralp2017foresight}
\bibfield{author}{\bibinfo{person}{\c{C}a\u{g}atay Demiralp}, \bibinfo{person}{Peter~J. Haas}, \bibinfo{person}{Srinivasan Parthasarathy}, {and} \bibinfo{person}{Tejaswini Pedapati}.} \bibinfo{year}{2017}\natexlab{}.
\newblock \showarticletitle{{Foresight: Recommending Visual Insights}}.
\newblock \bibinfo{journal}{\emph{Proc. VLDB Endow.}} \bibinfo{volume}{10}, \bibinfo{number}{12} (\bibinfo{date}{Aug.} \bibinfo{year}{2017}), \bibinfo{pages}{1937–1940}.
\newblock
\showISSN{2150-8097}
\urldef\tempurl%
\url{https://doi.org/10.14778/3137765.3137813}
\showDOI{\tempurl}


\bibitem[Duda and Hart(1974)]%
        {Duda1974PatternCA}
\bibfield{author}{\bibinfo{person}{Richard~O. Duda} {and} \bibinfo{person}{Peter~E. Hart}.} \bibinfo{year}{1974}\natexlab{}.
\newblock \showarticletitle{{Pattern Classification and Scene Analysis}}. In \bibinfo{booktitle}{\emph{{A Wiley-Interscience Publication}}}.
\newblock
\urldef\tempurl%
\url{https://api.semanticscholar.org/CorpusID:12946615}
\showURL{%
\tempurl}


\bibitem[Fan et~al\mbox{.}(2024)]%
        {fan2024understanding}
\bibfield{author}{\bibinfo{person}{Arlen Fan}, \bibinfo{person}{Fan Lei}, \bibinfo{person}{Michelle Mancenido}, \bibinfo{person}{Alan~M. Maceachren}, {and} \bibinfo{person}{Ross Maciejewski}.} \bibinfo{year}{2024}\natexlab{}.
\newblock \showarticletitle{Understanding Reader Takeaways in Thematic Maps Under Varying Text, Detail, and Spatial Autocorrelation}. In \bibinfo{booktitle}{\emph{Proceedings of the 2024 CHI Conference on Human Factors in Computing Systems}} (Honolulu, HI, USA) \emph{(\bibinfo{series}{CHI '24})}. \bibinfo{publisher}{Association for Computing Machinery}, \bibinfo{address}{New York, NY, USA}, Article \bibinfo{articleno}{213}, \bibinfo{numpages}{17}~pages.
\newblock
\showISBNx{9798400703300}
\urldef\tempurl%
\url{https://doi.org/10.1145/3613904.3642132}
\showDOI{\tempurl}


\bibitem[Gao et~al\mbox{.}(2015)]%
        {gao2015datatone}
\bibfield{author}{\bibinfo{person}{Tong Gao}, \bibinfo{person}{Mira Dontcheva}, \bibinfo{person}{Eytan Adar}, \bibinfo{person}{Zhicheng Liu}, {and} \bibinfo{person}{Karrie~G Karahalios}.} \bibinfo{year}{2015}\natexlab{}.
\newblock \showarticletitle{{Datatone: Managing Ambiguity in Natural Language Interfaces for Data Visualization}}. In \bibinfo{booktitle}{\emph{Proceedings of the 28th Annual ACM Symposium on User Interface Software \& Technology (UIST)}}. \bibinfo{pages}{489--500}.
\newblock


\bibitem[He et~al\mbox{.}(2024)]%
        {he2024leveraging}
\bibfield{author}{\bibinfo{person}{Yi He}, \bibinfo{person}{Shixiong Cao}, \bibinfo{person}{Yang Shi}, \bibinfo{person}{Qing Chen}, \bibinfo{person}{Ke Xu}, {and} \bibinfo{person}{Nan Cao}.} \bibinfo{year}{2024}\natexlab{}.
\newblock \showarticletitle{{Leveraging Large Models for Crafting Narrative Visualization: A Survey}}.
\newblock \bibinfo{journal}{\emph{arXiv preprint arXiv:2401.14010}} (\bibinfo{year}{2024}).
\newblock


\bibitem[Horvitz(1999)]%
        {horvitz1999}
\bibfield{author}{\bibinfo{person}{Eric Horvitz}.} \bibinfo{year}{1999}\natexlab{}.
\newblock \showarticletitle{{Principles of Mixed-Initiative User Interfaces}}. In \bibinfo{booktitle}{\emph{Proceedings of the SIGCHI Conference on Human Factors in Computing Systems}} (Pittsburgh, Pennsylvania, USA) \emph{(\bibinfo{series}{CHI '99})}. \bibinfo{publisher}{Association for Computing Machinery}, \bibinfo{address}{New York, NY, USA}, \bibinfo{pages}{159–166}.
\newblock
\showISBNx{0201485591}
\urldef\tempurl%
\url{https://doi.org/10.1145/302979.303030}
\showDOI{\tempurl}


\bibitem[Hsu et~al\mbox{.}(2021)]%
        {hsu2021scicap}
\bibfield{author}{\bibinfo{person}{Ting-Yao Hsu}, \bibinfo{person}{C~Lee Giles}, {and} \bibinfo{person}{Ting-Hao Huang}.} \bibinfo{year}{2021}\natexlab{}.
\newblock \showarticletitle{{S}ci{C}ap: Generating Captions for Scientific Figures}. In \bibinfo{booktitle}{\emph{Findings of the Association for Computational Linguistics: EMNLP 2021}}, \bibfield{editor}{\bibinfo{person}{Marie-Francine Moens}, \bibinfo{person}{Xuanjing Huang}, \bibinfo{person}{Lucia Specia}, {and} \bibinfo{person}{Scott Wen-tau Yih}} (Eds.). \bibinfo{publisher}{Association for Computational Linguistics}, \bibinfo{address}{Punta Cana, Dominican Republic}, \bibinfo{pages}{3258--3264}.
\newblock
\urldef\tempurl%
\url{https://doi.org/10.18653/v1/2021.findings-emnlp.277}
\showDOI{\tempurl}


\bibitem[Hsu et~al\mbox{.}(2024)]%
        {hsu2024scicapenter}
\bibfield{author}{\bibinfo{person}{Ting-Yao Hsu}, \bibinfo{person}{Chieh-Yang Huang}, \bibinfo{person}{Shih-Hong Huang}, \bibinfo{person}{Ryan Rossi}, \bibinfo{person}{Sungchul Kim}, \bibinfo{person}{Tong Yu}, \bibinfo{person}{C~Lee Giles}, {and} \bibinfo{person}{Ting-Hao~Kenneth Huang}.} \bibinfo{year}{2024}\natexlab{}.
\newblock \showarticletitle{SciCapenter: Supporting Caption Composition for Scientific Figures with Machine-Generated Captions and Ratings}. In \bibinfo{booktitle}{\emph{Extended Abstracts of the 2024 CHI Conference on Human Factors in Computing Systems}} \emph{(\bibinfo{series}{CHI EA '24})}. \bibinfo{publisher}{Association for Computing Machinery}, \bibinfo{address}{New York, NY, USA}, Article \bibinfo{articleno}{284}, \bibinfo{numpages}{9}~pages.
\newblock
\showISBNx{9798400703317}
\urldef\tempurl%
\url{https://doi.org/10.1145/3613905.3650738}
\showDOI{\tempurl}


\bibitem[Jung et~al\mbox{.}(2021)]%
        {jung2021communicating}
\bibfield{author}{\bibinfo{person}{Crescentia Jung}, \bibinfo{person}{Shubham Mehta}, \bibinfo{person}{Atharva Kulkarni}, \bibinfo{person}{Yuhang Zhao}, {and} \bibinfo{person}{Yea-Seul Kim}.} \bibinfo{year}{2021}\natexlab{}.
\newblock \showarticletitle{{Communicating Visualizations Without Visuals: Investigation of Visualization Alternative Text for People with Visual Impairments}}.
\newblock \bibinfo{journal}{\emph{{IEEE Transactions on Visualization and Computer Graphics}}} \bibinfo{volume}{28}, \bibinfo{number}{1} (\bibinfo{year}{2021}), \bibinfo{pages}{1095--1105}.
\newblock


\bibitem[Kim et~al\mbox{.}(2023a)]%
        {kim2023emphasischecker}
\bibfield{author}{\bibinfo{person}{Dae~Hyun Kim}, \bibinfo{person}{Seulgi Choi}, \bibinfo{person}{Juho Kim}, \bibinfo{person}{Vidya Setlur}, {and} \bibinfo{person}{Maneesh Agrawala}.} \bibinfo{year}{2023}\natexlab{a}.
\newblock \showarticletitle{EmphasisChecker A Tool for Guiding Chart and Caption Emphasis}.
\newblock \bibinfo{journal}{\emph{IEEE Transactions on Visualization and Computer Graphics}} (\bibinfo{year}{2023}).
\newblock


\bibitem[Kim et~al\mbox{.}(2021)]%
        {kim2021towards}
\bibfield{author}{\bibinfo{person}{Dae~Hyun Kim}, \bibinfo{person}{Vidya Setlur}, {and} \bibinfo{person}{Maneesh Agrawala}.} \bibinfo{year}{2021}\natexlab{}.
\newblock \showarticletitle{{Towards Understanding How Readers Integrate Charts and Captions: A Case Study with Line Charts}}. In \bibinfo{booktitle}{\emph{Proceedings of the 2021 CHI Conference on Human Factors in Computing Systems}}. \bibinfo{pages}{1--11}.
\newblock


\bibitem[Kim et~al\mbox{.}(2023b)]%
        {kim2023explain}
\bibfield{author}{\bibinfo{person}{Gyeongri Kim}, \bibinfo{person}{Jiho Kim}, {and} \bibinfo{person}{Yea-Seul Kim}.} \bibinfo{year}{2023}\natexlab{b}.
\newblock \showarticletitle{``Explain What a Treemap is'': Exploratory Investigation of Strategies for Explaining Unfamiliar Chart to Blind and Low Vision Users}. In \bibinfo{booktitle}{\emph{Proceedings of the 2023 CHI Conference on Human Factors in Computing Systems}}. \bibinfo{pages}{1--13}.
\newblock


\bibitem[Kim et~al\mbox{.}(2024)]%
        {kim2024datadive}
\bibfield{author}{\bibinfo{person}{Hyunwoo Kim}, \bibinfo{person}{Khanh~Duy Le}, \bibinfo{person}{Gionnieve Lim}, \bibinfo{person}{Dae~Hyun Kim}, \bibinfo{person}{Yoo~Jin Hong}, {and} \bibinfo{person}{Juho Kim}.} \bibinfo{year}{2024}\natexlab{}.
\newblock \showarticletitle{DataDive: Supporting Readers' Contextualization of Statistical Statements with Data Exploration}. In \bibinfo{booktitle}{\emph{Proceedings of the 29th International Conference on Intelligent User Interfaces}}. \bibinfo{pages}{623--639}.
\newblock


\bibitem[Kong et~al\mbox{.}(2018)]%
        {kong2018frames}
\bibfield{author}{\bibinfo{person}{Ha-Kyung Kong}, \bibinfo{person}{Zhicheng Liu}, {and} \bibinfo{person}{Karrie Karahalios}.} \bibinfo{year}{2018}\natexlab{}.
\newblock \showarticletitle{{Frames and Slants in Titles of Visualizations on Controversial Topics}}. In \bibinfo{booktitle}{\emph{Proceedings of the 2018 CHI Conference on Human Factors in Computing Systems}}. \bibinfo{pages}{1--12}.
\newblock


\bibitem[Kong et~al\mbox{.}(2019)]%
        {kong2019trust}
\bibfield{author}{\bibinfo{person}{Ha-Kyung Kong}, \bibinfo{person}{Zhicheng Liu}, {and} \bibinfo{person}{Karrie Karahalios}.} \bibinfo{year}{2019}\natexlab{}.
\newblock \showarticletitle{{Trust and Recall of Information Across Varying Degrees of Title-visualization Misalignment}}. In \bibinfo{booktitle}{\emph{Proceedings of the 2019 CHI Conference on Human Factors in Computing Systems}}. \bibinfo{pages}{1--13}.
\newblock


\bibitem[Lam et~al\mbox{.}(2012)]%
        {lam2012}
\bibfield{author}{\bibinfo{person}{Heidi Lam}, \bibinfo{person}{Enrico Bertini}, \bibinfo{person}{Petra Isenberg}, \bibinfo{person}{Catherine Plaisant}, {and} \bibinfo{person}{Sheelagh Carpendale}.} \bibinfo{year}{2012}\natexlab{}.
\newblock \showarticletitle{{Empirical Studies in Information Visualization: Seven Scenarios}}.
\newblock \bibinfo{journal}{\emph{IEEE Transactions on Visualization and Computer Graphics}} \bibinfo{volume}{18}, \bibinfo{number}{9} (\bibinfo{year}{2012}), \bibinfo{pages}{1520--1536}.
\newblock
\urldef\tempurl%
\url{https://doi.org/10.1109/TVCG.2011.279}
\showDOI{\tempurl}


\bibitem[Latif et~al\mbox{.}(2021)]%
        {latif2021kori}
\bibfield{author}{\bibinfo{person}{Shahid Latif}, \bibinfo{person}{Zheng Zhou}, \bibinfo{person}{Yoon Kim}, \bibinfo{person}{Fabian Beck}, {and} \bibinfo{person}{Nam~Wook Kim}.} \bibinfo{year}{2021}\natexlab{}.
\newblock \showarticletitle{{Kori: Interactive Synthesis of Text and Charts in Data Documents}}.
\newblock \bibinfo{journal}{\emph{IEEE Transactions on Visualization and Computer Graphics}} \bibinfo{volume}{28}, \bibinfo{number}{1} (\bibinfo{year}{2021}), \bibinfo{pages}{184--194}.
\newblock


\bibitem[Lin et~al\mbox{.}(2023)]%
        {lin2023inksight}
\bibfield{author}{\bibinfo{person}{Yanna Lin}, \bibinfo{person}{Haotian Li}, \bibinfo{person}{Leni Yang}, \bibinfo{person}{Aoyu Wu}, {and} \bibinfo{person}{Huamin Qu}.} \bibinfo{year}{2023}\natexlab{}.
\newblock \showarticletitle{\new{Inksight: Leveraging Sketch Interaction for Documenting Chart Findings in Computational Notebooks}}.
\newblock \bibinfo{journal}{\emph{IEEE Transactions on Visualization and Computer Graphics}} (\bibinfo{year}{2023}).
\newblock


\bibitem[Liu et~al\mbox{.}(2023)]%
        {liu2023autotitle}
\bibfield{author}{\bibinfo{person}{Can Liu}, \bibinfo{person}{Yuhan Guo}, {and} \bibinfo{person}{Xiaoru Yuan}.} \bibinfo{year}{2023}\natexlab{}.
\newblock \showarticletitle{{AutoTitle: An Interactive Title Generator for Visualizations}}.
\newblock \bibinfo{journal}{\emph{IEEE Transactions on Visualization and Computer Graphics}} (\bibinfo{year}{2023}).
\newblock


\bibitem[Liu et~al\mbox{.}(2020)]%
        {liu2020autocaption}
\bibfield{author}{\bibinfo{person}{Can Liu}, \bibinfo{person}{Liwenhan Xie}, \bibinfo{person}{Yun Han}, \bibinfo{person}{Datong Wei}, {and} \bibinfo{person}{Xiaoru Yuan}.} \bibinfo{year}{2020}\natexlab{}.
\newblock \showarticletitle{{AutoCaption: An Approach to Generate Natural Language Description from Visualization Automatically}}. In \bibinfo{booktitle}{\emph{2020 IEEE Pacific Visualization Symposium (PacificVis)}}. IEEE, \bibinfo{pages}{191--195}.
\newblock


\bibitem[Louis and Nenkova(2014)]%
        {louis2014}
\bibfield{author}{\bibinfo{person}{Annie Louis} {and} \bibinfo{person}{Ani Nenkova}.} \bibinfo{year}{2014}\natexlab{}.
\newblock \showarticletitle{{Verbose, Laconic or Just Right: A Simple Computational Model of Content Appropriateness under Length Constraints}}.
\newblock \bibinfo{journal}{\emph{14th Conference of the European Chapter of the Association for Computational Linguistics 2014, EACL 2014}}.
\newblock
\urldef\tempurl%
\url{https://doi.org/10.3115/v1/E14-1067}
\showDOI{\tempurl}


\bibitem[Lundgard and Satyanarayan(2021)]%
        {lundgard2021accessible}
\bibfield{author}{\bibinfo{person}{Alan Lundgard} {and} \bibinfo{person}{Arvind Satyanarayan}.} \bibinfo{year}{2021}\natexlab{}.
\newblock \showarticletitle{{Accessible Visualization via Natural Language Descriptions: A Four-level Model of Semantic Content}}.
\newblock \bibinfo{journal}{\emph{IEEE Transactions on Visualization and Computer Graphics}} \bibinfo{volume}{28}, \bibinfo{number}{1} (\bibinfo{year}{2021}), \bibinfo{pages}{1073--1083}.
\newblock


\bibitem[Manelski and Krulee(1965)]%
        {manelski-krulee-1965-heuristic}
\bibfield{author}{\bibinfo{person}{Denis~M. Manelski} {and} \bibinfo{person}{Gilbert~K. Krulee}.} \bibinfo{year}{1965}\natexlab{}.
\newblock \showarticletitle{{A Heuristic Approach to Natural Language Processing}}. In \bibinfo{booktitle}{\emph{COLING 1965}}.
\newblock
\urldef\tempurl%
\url{https://aclanthology.org/C65-1018}
\showURL{%
\tempurl}


\bibitem[Narechania et~al\mbox{.}(2020)]%
        {narechania2020nl4dv}
\bibfield{author}{\bibinfo{person}{Arpit Narechania}, \bibinfo{person}{Arjun Srinivasan}, {and} \bibinfo{person}{John Stasko}.} \bibinfo{year}{2020}\natexlab{}.
\newblock \showarticletitle{{NL4DV: A Toolkit for Generating Analytic Specifications for Data Visualization from Natural Language Queries}}.
\newblock \bibinfo{journal}{\emph{IEEE Transactions on Visualization and Computer Graphics}} \bibinfo{volume}{27}, \bibinfo{number}{2} (\bibinfo{year}{2020}), \bibinfo{pages}{369--379}.
\newblock


\bibitem[Obeid and Hoque(2020)]%
        {obeid2020chart}
\bibfield{author}{\bibinfo{person}{Jason Obeid} {and} \bibinfo{person}{Enamul Hoque}.} \bibinfo{year}{2020}\natexlab{}.
\newblock \showarticletitle{Chart-to-Text: Generating Natural Language Descriptions for Charts by Adapting the Transformer Model}. In \bibinfo{booktitle}{\emph{International Conference on Natural Language Generation}}.
\newblock
\urldef\tempurl%
\url{https://api.semanticscholar.org/CorpusID:224704949}
\showURL{%
\tempurl}


\bibitem[Ottley et~al\mbox{.}(2019)]%
        {ottley2019curious}
\bibfield{author}{\bibinfo{person}{Alvitta Ottley}, \bibinfo{person}{Aleksandra Kaszowska}, \bibinfo{person}{R~Jordan Crouser}, {and} \bibinfo{person}{Evan~M Peck}.} \bibinfo{year}{2019}\natexlab{}.
\newblock \showarticletitle{{The Curious Case of Combining Text and Visualization}}.
\newblock \bibinfo{journal}{\emph{EuroVis 2019-Short Papers}} (\bibinfo{year}{2019}).
\newblock


\bibitem[Purich et~al\mbox{.}(2023)]%
        {purich2023toward}
\bibfield{author}{\bibinfo{person}{Joanna Purich}, \bibinfo{person}{Arjun Srinivasan}, \bibinfo{person}{Michael Correll}, \bibinfo{person}{Leilani Battle}, \bibinfo{person}{Vidya Setlur}, {and} \bibinfo{person}{Anamaria Crisan}.} \bibinfo{year}{2023}\natexlab{}.
\newblock \showarticletitle{\new{Toward a Scalable Census of Dashboard Designs in the Wild: A Case Study with Tableau Public}}.
\newblock \bibinfo{journal}{\emph{arXiv preprint arXiv:2306.16513}} (\bibinfo{year}{2023}).
\newblock


\bibitem[Quadri et~al\mbox{.}(2024)]%
        {quadri2024you}
\bibfield{author}{\bibinfo{person}{Ghulam~Jilani Quadri}, \bibinfo{person}{Arran~Zeyu Wang}, \bibinfo{person}{Zhehao Wang}, \bibinfo{person}{Jennifer Adorno}, \bibinfo{person}{Paul Rosen}, {and} \bibinfo{person}{Danielle~Albers Szafir}.} \bibinfo{year}{2024}\natexlab{}.
\newblock \showarticletitle{Do You See What I See? A Qualitative Study Eliciting High-Level Visualization Comprehension}. In \bibinfo{booktitle}{\emph{Proceedings of the 2024 CHI Conference on Human Factors in Computing Systems}} (Honolulu, HI, USA) \emph{(\bibinfo{series}{CHI '24})}. \bibinfo{publisher}{Association for Computing Machinery}, \bibinfo{address}{New York, NY, USA}, Article \bibinfo{articleno}{204}, \bibinfo{numpages}{26}~pages.
\newblock
\showISBNx{9798400703300}
\urldef\tempurl%
\url{https://doi.org/10.1145/3613904.3642813}
\showDOI{\tempurl}


\bibitem[Saket et~al\mbox{.}(2018)]%
        {saket2018task}
\bibfield{author}{\bibinfo{person}{Bahador Saket}, \bibinfo{person}{Alex Endert}, {and} \bibinfo{person}{{\c{C}}a{\u{g}}atay Demiralp}.} \bibinfo{year}{2018}\natexlab{}.
\newblock \showarticletitle{{Task-Based Effectiveness of Basic Visualizations}}.
\newblock \bibinfo{journal}{\emph{IEEE Transactions on Visualization and Computer Graphics}} \bibinfo{volume}{25}, \bibinfo{number}{7} (\bibinfo{year}{2018}), \bibinfo{pages}{2505--2512}.
\newblock


\bibitem[Sanh et~al\mbox{.}(2019)]%
        {sanh2019distilbert}
\bibfield{author}{\bibinfo{person}{Victor Sanh}, \bibinfo{person}{Lysandre Debut}, \bibinfo{person}{Julien Chaumond}, {and} \bibinfo{person}{Thomas Wolf}.} \bibinfo{year}{2019}\natexlab{}.
\newblock \showarticletitle{{DistilBERT, A Distilled Version of BERT: Smaller, Faster, Cheaper and Lighter}}.
\newblock \bibinfo{journal}{\emph{arXiv preprint arXiv:1910.01108}} (\bibinfo{year}{2019}).
\newblock


\bibitem[Sarikaya et~al\mbox{.}(2018)]%
        {sarikaya2018we}
\bibfield{author}{\bibinfo{person}{Alper Sarikaya}, \bibinfo{person}{Michael Correll}, \bibinfo{person}{Lyn Bartram}, \bibinfo{person}{Melanie Tory}, {and} \bibinfo{person}{Danyel Fisher}.} \bibinfo{year}{2018}\natexlab{}.
\newblock \showarticletitle{What Do We Talk About When We Talk About Dashboards?}
\newblock \bibinfo{journal}{\emph{IEEE Transactions on Visualization and Computer Graphics}} \bibinfo{volume}{25}, \bibinfo{number}{1} (\bibinfo{year}{2018}), \bibinfo{pages}{682--692}.
\newblock


\bibitem[Satyanarayan et~al\mbox{.}(2016)]%
        {satyanarayan2016vega}
\bibfield{author}{\bibinfo{person}{Arvind Satyanarayan}, \bibinfo{person}{Dominik Moritz}, \bibinfo{person}{Kanit Wongsuphasawat}, {and} \bibinfo{person}{Jeffrey Heer}.} \bibinfo{year}{2016}\natexlab{}.
\newblock \showarticletitle{{Vega-lite: A Grammar of Interactive Graphics}}.
\newblock \bibinfo{journal}{\emph{IEEE Transactions on Visualization and Computer Graphics}} \bibinfo{volume}{23}, \bibinfo{number}{1} (\bibinfo{year}{2016}), \bibinfo{pages}{341--350}.
\newblock


\bibitem[Schulz et~al\mbox{.}(2013)]%
        {schulz2013design}
\bibfield{author}{\bibinfo{person}{Hans-J{\"o}rg Schulz}, \bibinfo{person}{Thomas Nocke}, \bibinfo{person}{Magnus Heitzler}, {and} \bibinfo{person}{Heidrun Schumann}.} \bibinfo{year}{2013}\natexlab{}.
\newblock \showarticletitle{{A Design Space of Visualization Tasks}}.
\newblock \bibinfo{journal}{\emph{IEEE Transactions on Visualization and Computer Graphics}} \bibinfo{volume}{19}, \bibinfo{number}{12} (\bibinfo{year}{2013}), \bibinfo{pages}{2366--2375}.
\newblock


\bibitem[Segel and Heer(2010)]%
        {segel2010narrative}
\bibfield{author}{\bibinfo{person}{Edward Segel} {and} \bibinfo{person}{Jeffrey Heer}.} \bibinfo{year}{2010}\natexlab{}.
\newblock \showarticletitle{Narrative {V}isualization: {T}elling {S}tories with {D}ata}.
\newblock \bibinfo{journal}{\emph{IEEE Transactions on Visualization and Computer Graphics}} \bibinfo{volume}{16}, \bibinfo{number}{6} (\bibinfo{year}{2010}), \bibinfo{pages}{1139--1148}.
\newblock


\bibitem[Setlur et~al\mbox{.}(2016)]%
        {setlur2016eviza}
\bibfield{author}{\bibinfo{person}{Vidya Setlur}, \bibinfo{person}{Sarah~E Battersby}, \bibinfo{person}{Melanie Tory}, \bibinfo{person}{Rich Gossweiler}, {and} \bibinfo{person}{Angel~X Chang}.} \bibinfo{year}{2016}\natexlab{}.
\newblock \showarticletitle{{Eviza: A Natural Language Interface for Visual Analysis}}. In \bibinfo{booktitle}{\emph{Proceedings of the 29th Annual Symposium on User Interface Software and Technology (UIST)}}. \bibinfo{pages}{365--377}.
\newblock


\bibitem[Shen et~al\mbox{.}(2024)]%
        {shen2024data}
\bibfield{author}{\bibinfo{person}{Leixian Shen}, \bibinfo{person}{Haotian Li}, \bibinfo{person}{Yun Wang}, \bibinfo{person}{Tianqi Luo}, \bibinfo{person}{Yuyu Luo}, {and} \bibinfo{person}{Huamin Qu}.} \bibinfo{year}{2024}\natexlab{}.
\newblock \showarticletitle{\new{Data Playwright: Authoring Data Videos with Annotated Narration}}.
\newblock \bibinfo{journal}{\emph{IEEE Transactions on Visualization and Computer Graphics}} (\bibinfo{year}{2024}).
\newblock


\bibitem[Shneiderman(2003)]%
        {shneiderman2003eyes}
\bibfield{author}{\bibinfo{person}{Ben Shneiderman}.} \bibinfo{year}{2003}\natexlab{}.
\newblock \showarticletitle{{The Eyes Have It: A Task by Data Type Taxonomy for Information Visualizations}}.
\newblock In \bibinfo{booktitle}{\emph{The Craft of Information Visualization}}. \bibinfo{publisher}{Elsevier}, \bibinfo{pages}{364--371}.
\newblock


\bibitem[Shneiderman and Plaisant(2006)]%
        {shneiderman2006strategies}
\bibfield{author}{\bibinfo{person}{Ben Shneiderman} {and} \bibinfo{person}{Catherine Plaisant}.} \bibinfo{year}{2006}\natexlab{}.
\newblock \showarticletitle{Strategies for Evaluating Information Visualization Tools: Multi-Dimensional In-Depth Long-Term Case Studies}. In \bibinfo{booktitle}{\emph{Proceedings of the 2006 AVI Workshop on BEyond Time and Errors: Novel Evaluation Methods for Information Visualization}}. \bibinfo{pages}{1--7}.
\newblock


\bibitem[Singh et~al\mbox{.}(2024)]%
        {singh2024figura11y}
\bibfield{author}{\bibinfo{person}{Nikhil Singh}, \bibinfo{person}{Lucy~Lu Wang}, {and} \bibinfo{person}{Jonathan Bragg}.} \bibinfo{year}{2024}\natexlab{}.
\newblock \showarticletitle{FigurA11y: AI Assistance for Writing Scientific Alt Text}. In \bibinfo{booktitle}{\emph{Proceedings of the 29th International Conference on Intelligent User Interfaces}}. \bibinfo{pages}{886--906}.
\newblock


\bibitem[Srinivasan et~al\mbox{.}(2018)]%
        {srinivasan2018augmenting}
\bibfield{author}{\bibinfo{person}{Arjun Srinivasan}, \bibinfo{person}{Steven~M Drucker}, \bibinfo{person}{Alex Endert}, {and} \bibinfo{person}{John Stasko}.} \bibinfo{year}{2018}\natexlab{}.
\newblock \showarticletitle{{Augmenting Visualizations with Interactive Data Facts to Facilitate Interpretation and Communication}}.
\newblock \bibinfo{journal}{\emph{IEEE Transactions on Visualization and Computer Graphics}} \bibinfo{volume}{25}, \bibinfo{number}{1} (\bibinfo{year}{2018}), \bibinfo{pages}{672--681}.
\newblock


\bibitem[Srinivasan and Setlur(2021)]%
        {snowy2021}
\bibfield{author}{\bibinfo{person}{Arjun Srinivasan} {and} \bibinfo{person}{Vidya Setlur}.} \bibinfo{year}{2021}\natexlab{}.
\newblock \showarticletitle{{Snowy: Recommending Utterances for Conversational Visual Analysis}}. In \bibinfo{booktitle}{\emph{The 34th Annual ACM Symposium on User Interface Software and Technology (UIST)}} (Virtual Event, USA) \emph{(\bibinfo{series}{UIST '21})}. \bibinfo{publisher}{Association for Computing Machinery}, \bibinfo{address}{New York, NY, USA}, \bibinfo{pages}{864–880}.
\newblock
\showISBNx{9781450386357}
\urldef\tempurl%
\url{https://doi.org/10.1145/3472749.3474792}
\showDOI{\tempurl}


\bibitem[Stokes et~al\mbox{.}(2023)]%
        {stokes2023role}
\bibfield{author}{\bibinfo{person}{Chase Stokes}, \bibinfo{person}{Cindy~Xiong Bearfield}, {and} \bibinfo{person}{Marti~A Hearst}.} \bibinfo{year}{2023}\natexlab{}.
\newblock \showarticletitle{{The Role of Text in Visualizations: How Annotations Shape Perceptions of Bias and Influence Predictions}}.
\newblock \bibinfo{journal}{\emph{IEEE Transactions on Visualization and Computer Graphics}} (\bibinfo{year}{2023}).
\newblock


\bibitem[Stokes and Hearst(2021)]%
        {stokesgive}
\bibfield{author}{\bibinfo{person}{Chase Stokes} {and} \bibinfo{person}{Marti~A Hearst}.} \bibinfo{year}{2021}\natexlab{}.
\newblock \showarticletitle{{Give Text A Chance: Advocating for Equal Consideration for Language and Visualization}}. In \bibinfo{booktitle}{\emph{{NL VIZ: Workshop on Exploring Opportunities and Challenges for Natural Language Techniques to Support Visual Analysis}}}. IEEE.
\newblock


\bibitem[Stokes et~al\mbox{.}(2022)]%
        {stokes2022striking}
\bibfield{author}{\bibinfo{person}{Chase Stokes}, \bibinfo{person}{Vidya Setlur}, \bibinfo{person}{Bridget Cogley}, \bibinfo{person}{Arvind Satyanarayan}, {and} \bibinfo{person}{Marti~A Hearst}.} \bibinfo{year}{2022}\natexlab{}.
\newblock \showarticletitle{{Striking a Balance: Reader Takeaways and Preferences when Integrating Text and Charts}}.
\newblock \bibinfo{journal}{\emph{IEEE Transactions on Visualization and Computer Graphics}} \bibinfo{volume}{29}, \bibinfo{number}{1} (\bibinfo{year}{2022}), \bibinfo{pages}{1233--1243}.
\newblock


\bibitem[Stolte et~al\mbox{.}(2008)]%
        {polaris}
\bibfield{author}{\bibinfo{person}{Chris Stolte}, \bibinfo{person}{Diane Tang}, {and} \bibinfo{person}{Pat Hanrahan}.} \bibinfo{year}{2008}\natexlab{}.
\newblock \showarticletitle{{Polaris: A System for Query, Analysis, and Visualization of Multidimensional Databases}}.
\newblock \bibinfo{journal}{\emph{Commun. ACM}} \bibinfo{volume}{51}, \bibinfo{number}{11} (\bibinfo{date}{Nov} \bibinfo{year}{2008}), \bibinfo{pages}{75–84}.
\newblock
\showISSN{0001-0782}
\urldef\tempurl%
\url{https://doi.org/10.1145/1400214.1400234}
\showDOI{\tempurl}


\bibitem[Sultanum et~al\mbox{.}(2021)]%
        {sultanum2021}
\bibfield{author}{\bibinfo{person}{Nicole Sultanum}, \bibinfo{person}{Fanny Chevalier}, \bibinfo{person}{Zoya Bylinskii}, {and} \bibinfo{person}{Zhicheng Liu}.} \bibinfo{year}{2021}\natexlab{}.
\newblock \showarticletitle{{Leveraging Text-Chart Links to Support Authoring of Data-Driven Articles with VizFlow}}. In \bibinfo{booktitle}{\emph{Proceedings of the 2021 CHI Conference on Human Factors in Computing Systems}} (Yokohama, Japan) \emph{(\bibinfo{series}{CHI '21})}. \bibinfo{publisher}{Association for Computing Machinery}, \bibinfo{address}{New York, NY, USA}, Article \bibinfo{articleno}{16}, \bibinfo{numpages}{17}~pages.
\newblock
\showISBNx{9781450380966}
\urldef\tempurl%
\url{https://doi.org/10.1145/3411764.3445354}
\showDOI{\tempurl}


\bibitem[Sultanum and Srinivasan(2023)]%
        {sultanum2023datatales}
\bibfield{author}{\bibinfo{person}{Nicole Sultanum} {and} \bibinfo{person}{Arjun Srinivasan}.} \bibinfo{year}{2023}\natexlab{}.
\newblock \showarticletitle{{Datatales: Investigating the Use of Large Language Models for Authoring Data-driven Articles}}. In \bibinfo{booktitle}{\emph{2023 IEEE Visualization and Visual Analytics (VIS)}}. IEEE, \bibinfo{pages}{231--235}.
\newblock


\bibitem[Tang et~al\mbox{.}(2023)]%
        {tang2023vistext}
\bibfield{author}{\bibinfo{person}{Benny~J Tang}, \bibinfo{person}{Angie Boggust}, {and} \bibinfo{person}{Arvind Satyanarayan}.} \bibinfo{year}{2023}\natexlab{}.
\newblock \showarticletitle{{Vistext: A Benchmark for Semantically Rich Chart Captioning}}.
\newblock \bibinfo{journal}{\emph{arXiv preprint arXiv:2307.05356}} (\bibinfo{year}{2023}).
\newblock


\bibitem[Walker and Duncan(1967)]%
        {strother1967}
\bibfield{author}{\bibinfo{person}{Strother~H. Walker} {and} \bibinfo{person}{David~B. Duncan}.} \bibinfo{year}{1967}\natexlab{}.
\newblock \showarticletitle{{Estimation of the Probability of an Event as a Function of Several Independent Variables}}.
\newblock \bibinfo{journal}{\emph{Biometrika}} \bibinfo{volume}{54}, \bibinfo{number}{1/2} (\bibinfo{year}{1967}), \bibinfo{pages}{167--179}.
\newblock
\showISSN{00063444}
\urldef\tempurl%
\url{http://www.jstor.org/stable/2333860}
\showURL{%
\tempurl}


\bibitem[Wang et~al\mbox{.}(2022)]%
        {wang2022towards}
\bibfield{author}{\bibinfo{person}{Yun Wang}, \bibinfo{person}{Zhitao Hou}, \bibinfo{person}{Leixian Shen}, \bibinfo{person}{Tongshuang Wu}, \bibinfo{person}{Jiaqi Wang}, \bibinfo{person}{He Huang}, \bibinfo{person}{Haidong Zhang}, {and} \bibinfo{person}{Dongmei Zhang}.} \bibinfo{year}{2022}\natexlab{}.
\newblock \showarticletitle{\new{Towards Natural Language-based Visualization Authoring}}.
\newblock \bibinfo{journal}{\emph{IEEE Transactions on Visualization and Computer Graphics}} \bibinfo{volume}{29}, \bibinfo{number}{1} (\bibinfo{year}{2022}), \bibinfo{pages}{1222--1232}.
\newblock


\bibitem[Wang et~al\mbox{.}(2019)]%
        {wang2019datashot}
\bibfield{author}{\bibinfo{person}{Yun Wang}, \bibinfo{person}{Zhida Sun}, \bibinfo{person}{Haidong Zhang}, \bibinfo{person}{Weiwei Cui}, \bibinfo{person}{Ke Xu}, \bibinfo{person}{Xiaojuan Ma}, {and} \bibinfo{person}{Dongmei Zhang}.} \bibinfo{year}{2019}\natexlab{}.
\newblock \showarticletitle{{Datashot: Automatic Generation of Fact Sheets from Tabular Data}}.
\newblock \bibinfo{journal}{\emph{IEEE Transactions on Visualization and Computer Graphics}} \bibinfo{volume}{26}, \bibinfo{number}{1} (\bibinfo{year}{2019}), \bibinfo{pages}{895--905}.
\newblock


\bibitem[Wu and Palmer(1994)]%
        {wu1994verb}
\bibfield{author}{\bibinfo{person}{Zhibiao Wu} {and} \bibinfo{person}{Martha Palmer}.} \bibinfo{year}{1994}\natexlab{}.
\newblock \showarticletitle{Verbs Semantics and Lexical Selection}. In \bibinfo{booktitle}{\emph{Proceedings of the 32nd Annual Meeting on Association for Computational Linguistics}} (Las Cruces, New Mexico) \emph{(\bibinfo{series}{ACL '94})}. \bibinfo{publisher}{Association for Computational Linguistics}, \bibinfo{address}{USA}, \bibinfo{pages}{133–138}.
\newblock
\urldef\tempurl%
\url{https://doi.org/10.3115/981732.981751}
\showDOI{\tempurl}


\bibitem[Yaneva et~al\mbox{.}(2015)]%
        {yaneva2015accessible}
\bibfield{author}{\bibinfo{person}{Victoria Yaneva}, \bibinfo{person}{Irina Temnikova}, {and} \bibinfo{person}{Ruslan Mitkov}.} \bibinfo{year}{2015}\natexlab{}.
\newblock \showarticletitle{{Accessible Texts for Autism: An Eye-Tracking Study}}. In \bibinfo{booktitle}{\emph{Proceedings of the 17th International ACM SIGACCESS Conference on Computers \& Accessibility}}. \bibinfo{pages}{49--57}.
\newblock


\bibitem[Yujian and Bo(2007)]%
        {yujian2007normalized}
\bibfield{author}{\bibinfo{person}{Li Yujian} {and} \bibinfo{person}{Liu Bo}.} \bibinfo{year}{2007}\natexlab{}.
\newblock \showarticletitle{{A Normalized Levenshtein Distance Metric}}.
\newblock \bibinfo{journal}{\emph{IEEE Transactions on Pattern Analysis and Machine Intelligence}} \bibinfo{volume}{29}, \bibinfo{number}{6} (\bibinfo{year}{2007}), \bibinfo{pages}{1091--1095}.
\newblock


\end{thebibliography}

\end{document}